\documentclass[fleqn,usenatbib]{mnras} 

\usepackage{graphicx}	
\usepackage{amsmath}	
\usepackage{amssymb}	

\usepackage{multirow}

\newcommand{\kms}{km\,s$^{-1}$} 

\title{Composite hot-subdwarf binaries -- I. The spectroscopically confirmed sdB sample.}

\author[J. Vos et al.]{
Joris Vos$^{1}$\thanks{E-mail: joris.vos@uv.cl, Based on observations collected at the European Southern Observatory, Chile under programme IDs 088.D-0364, 093.D-0629, 096.D-0180, 097.D-0110, 098.D-0018 and 099.D-0014},
P\'{e}ter N\'{e}meth$^{2,3}$,
Maja Vu\u{c}kovi\'{c}$^{1}$,
Roy \O{}stensen$^{4}$
and
\newauthor
Steven Parsons$^{5}$
\\
$^{1}$Instituto de F\'{\i}sica y Astronom\'{\i}a, Universidad de Valparaiso, Gran Breta\~{n}a 1111, Playa Ancha, Valpara\'{\i}so
2360102, Chile\\
$^{2}$Astroserver.org, 8533 Malomsok, Hungary\\
$^{3}$Dr. Karl-Remeis-Observatory \& ECAP, Astronomical Institute, F.-A.-U. Erlangen-N\"{u}rnberg, 96049 Bamberg, Germany\\
$^{4}$Department of Physics, Astronomy, and Materials Science, Missouri State University, Springfield, MO 65804, USA\\
$^{5}$Department of Physics and Astronomy, University of Sheffield, Sheffield, S3 7RH, UK
}

\date{Accepted XXX. Received YYY; in original form ZZZ}

\pubyear{2015}

\begin{document}
\label{firstpage}
\pagerange{\pageref{firstpage}--\pageref{lastpage}}
\maketitle

\begin{abstract}
Hot subdwarf-B (sdB) stars in long-period binaries are found to be on eccentric orbits, even though current binary-evolution theory predicts these objects to be circularised before the onset of Roche-lobe overflow (RLOF). To increase our understanding of binary interaction processes during the RLOF phase, we started a long term observing campain to study wide sdB binaries. In this article we present a composite-binary-sdB sample, and the results of the spectral analysis of 9 such systems. The grid search in stellar parameters (GSSP) code is used to derive atmospheric parameters for the cool companions. To cross-check our results and also characterize the hot subdwarfs we used the independent {\sc XTgrid} code, which employs {\sc Tlusty} non-local thermodynamic equilibrium models to derive atmospheric parameters for the sdB component and {\sc Phoenix} synthetic spectra for the cool companions. The independent GSSP and {\sc XTgrid} codes are found to show good agreement for three test systems that have atmospheric parameters available in the literature. Based on the rotational velocity of the companions, an estimate for the mass accreted during the RLOF phase and the miminum duration of that phase is made. It is found that the mass transfer to the companion is minimal during the subdwarf formation. 
\end{abstract}




\section{Introduction}
Hot subdwarf-B (sdB) stars are core-helium-burning stars with a very thin hydrogen envelope (M$_{\rm{H}}$ $<$ 0.02 $M_{\odot}$), and a mass close to the core-helium-flash mass $\sim$ 0.47 $M_{\odot}$ \citep{Saffer1994, Brassard2001}. These hot subdwarfs are found in all galactic populations, and they are the main source for the UV-upturn in early-type galaxies \citep{Green1986, Greggio1990, Brown1997}. Furthermore, their photospheric chemical composition is governed by diffusion processes causing strong He-depletion and other chemical peculiarities \citep{Heber1998}. 
The formation of these extreme-horizontal-branch (EHB) objects is still puzzling. To form an sdB star, its progenitor needs to lose its hydrogen envelope almost completely before reaching the tip of the red-giant branch (RGB), so that the He-core ignites while the remaining hydrogen envelope is not massive enough to sustain hydrogen-shell burning. A variety of possible formation channels have been proposed. 
Currently, there is a consensus that sdB stars are only formed in binaries. Several evolutionary channels have been proposed where binary-interaction physics plays a major role. Close binary systems can be formed in a common-envelope (CE) ejection channel \citep{Paczynski1976}, while stable Roche-lobe overflow (RLOF) can produce wide sdB binaries \citep{Han2000, Han2002}. An alternative formation channel forming a single sdB star is the double white-dwarf (WD) merger, where a pair of low-mass He-core white dwarfs spiral in and merge to form a single sdB star \citep{Webbink1984}.

\citet{Han2002, Han2003} addressed these three binary-formation mechanisms, and performed binary-population-synthesis (BPS) studies for two kinds of CE-ejection channels, two possible stable-RLOF channels and the WD-merger channel. The CE-ejection channels produce close binaries with periods of $P_{\rm{orb}}$ = 0.1 -- 10 d, and main-sequence (MS) or white-dwarf (WD) companions. The sdB binaries formed through stable RLOF have orbital periods ranging from 10 to 500 days, and MS companions. \citet{Chen2013} revisited the RLOF models of \citet{Han2003} with a more sophisticated treatment of angular-momentum loss. When including atmospheric RLOF, these revised models can reach orbital periods as long as $\sim$1600\ d. Finally, The WD-merger channel can lead to single sdB stars with a higher mass, up to 0.65 $M_{\odot}$. A detailed review of hot subdwarf stars is given by \citet{Heber2016}.

Many observational studies have focused on short-period sdB binaries \citep{Koen1998, Maxted2000, Maxted2001, Morales2003, Napiwotzki2004, Copperwheat2011}, and 148 of these systems are currently known \citep{Kupfer2015, Kawka2015}. These observed short-period sdB binaries agree very well with the results of BPS studies. Currently, only nine long-period sdB binaries are known \citep{Green2001,Oestensen2011,Oestensen2012,Deca2012,Barlow2012,Vos2012,Vos2013,Vos2014}. Even though this is a small sample, their period-eccentricity distribution shows an unexpected trend. Eight of the nine systems have a significant eccentric orbit, and the entire sample shows a clear correlation of higher eccentricity with longer orbital period \cite{Vos2015}. 

\cite{Vos2015} showed that a combination of two eccentricity pumping mechanisms, phase dependent RLOF and a circumbinary (CB) disk, can explain the observed eccentricities of wide sdB binaries. However, these models are unable to reproduce the observed trend of higher eccentricities at longer orbital periods. The proposed mechanisms indicate even stronger eccentricity pumping at shorter orbital periods. The two eccentricity pumping processes are highly parameterised, and many of these parameters, e.g. the mass-loss parameters of \cite{Tauris2006}, the CB disk mass, the redistribution of mass in a CB disk and many more are currently unconstrained. 

To properly constrain the input parameters of the eccentricity pumping models, a statistically significant sample of well-studied wide sdB binaries is necessary. In addition to the orbital parameters, the spectroscopically determined properties of the companion star are very relevant to the exploration of any systematic relations between the model parameters and physical reality. The RLOF process in wide sdB binaries is normally considered to take place on a thermal timescale, since the sdB progenitor is an RGB star. However, the exact process including mass transfer between the members or mass loss from the system is not well understood. By studying the rotational velocity  of the companion, estimates about the accreted mass, and the minimal duration of the RLOF phase can be made. Such estimates are important input in simulations of the RLOF phase. 

A comprehensive catalog containing all currently known hot subdwarf stars was published by \citet{Geier2017}. This catalog contains 5613 sdO and sdB star, both single and in binary systems, which have been selected on photometric or spectroscopic criteria.

In this article we present a sample of spectroscopically confirmed composite sdB binaries (Sect.\,\ref{s:composite_sample}) created with the aim to study the binary interaction processes during RLOF. The ongoing long-term observing program to explore the orbits of composite sdB binaries is also described (Sect.\,\ref{s:observations}). Enough spectra have been obtained for nine systems to allow a detailed spectroscopic study of the companions of those wide sdB binaries. The methods employed as well as a discussion of their accuracy is given in Sect.\ref{s:methods}, while the atmospheric parameters are given in Sect.\,\ref{s:results} and mass accretion during the RLOF phase is discussed in Sect.\,\ref{s:accretion}. The article concludes with a summary and discussion.

\section{The composite sdB sample}\label{s:composite_sample}
The targets for the current study have been selected from an extensive sample of composite sdB binaries. This sample is created based on a literature search for spectroscopically confirmed composite sdB binaries. We required at least medium resolution spectroscopic observations to include a system as a confirmed composite binary. Most of these targets originated in the surveys of \citet{Green1986, Kilkenny1988, Bixler1991, Lamontagne2000, Aznar2001} and \citet{Reed2004}. This list of observed systems is supplemented with targets from the Edinburgh-Cape Survey \citep{Stobie1997}, and our own survey sample of UV excess targets from the Galaxy Evolution Explorer \citep[GALEX,][]{Martin2005} survey, where we selected those stars that had both UV and IR excess relative to visual magnitudes.

This sample was constructed independently from the hot-subdwarf catalog of \citet{Geier2017}, but can be seen as a subsample of this catalog. However, we have employed more strict selection criteria for a spectroscopic confirmation of the binary systems, rejecting systems that only have low resolution spectroscopy.Our catalog contains ten systems that are not included in the hot-subdwarf catalog of \citet{Geier2017}. Three systems from various surveys: Ton S 367, KUV 01542-071 and Balloon 82800003, and seven systems found among the UV excess targets from Galex: J00116-5423, J02151-7034, J02286-3625, J09000+0128, J17416+6414, J17534-5007 and J21402-3714.

In this sample we have only focused on systems containing an sdB component, and have excluded all systems with sdO components. The goals of this project is to create a test sample for stable RLOF evolution, of which wide sdB binaries are direct evolution products. Even though sdBs evolve into sdOs, they increase in radius, which can influence the orbital properties, making them less suitable for such a study. Furthermore, there are several known sdO+F/G systems, for example BD\,--11$^{\rm{o}}$162 \citep{Oestensen2012}, which have periods on the order of 100 days. These are most likely remnants of massive stars that lost their envelope on the RGB after non-degenerative He-core ignition. With such massive stars the envelope is much more tightly bound, and the parameters governing the mass transfer will be quite different.

The complete list of confirmed composite sdB binaries from the literature, supplemented with candidates that we have confirmed with our UVES observations as described in this paper, totals 148 systems, which are provided as an on-line VizieR catalogue. In Table\,\ref{tb:composite_sdbs}, just the first 10 lines of the complete table are given to show the structure. For each system the coordinates, spectral type, and reference are given. The reference refers to the most accurate classification and is provided in the standard 19 digit shorthand used by ADS. When available GALEX FUV and NUV magnitudes from data reduction 7, as well as B, V, g, r and i magnitudes from the 9th data reduction of the AAVSO Photometric All-Sky Survey \citep[APASS,][]{Henden2016} and 2MASS J, H and Ks magnitudes \citep{Skrutskie2006} are provided.  

For the systems observed with UVES, the standard classification system for hot subdwarfs is used \citep[see i.e,][]{Moehler1990}. Subdwarfs with broad Balmer lines and weak \ion{He}{I} lines are classified as sdB, while if the \ion{He}{II} $\lambda$ 4686 \AA\ line is also visible, they are classified as sdOB. If the systems were analysed with GSSP \citep{Tkachenko2015} or {\sc XTgrid} \citep{Nemeth2012}, the exact spectral type of the companion is given, otherwise they are indicated as MS.
The systems named Jnnnnn$\pm$nnnn are GALEX targets where the J number is shorthand for the full GALEX identification.

\begin{figure}
    \includegraphics{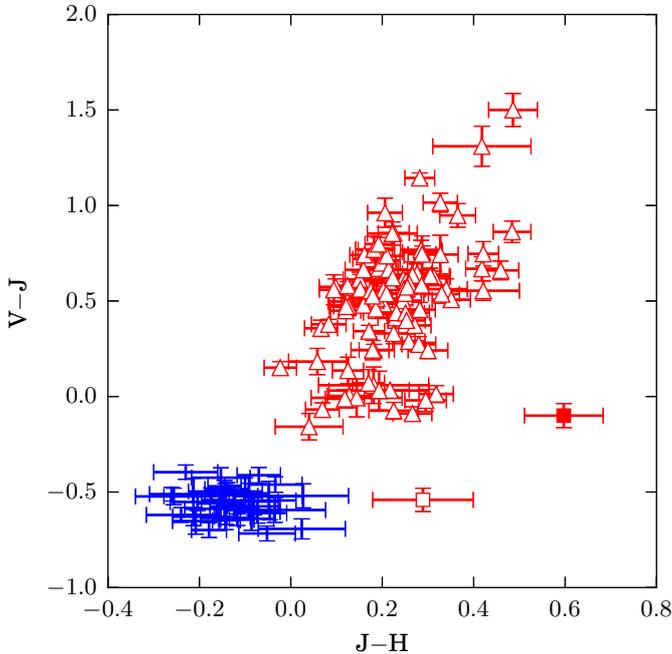}
    \caption{Colour-colour diagram of the composite sdB sample of 2MASS J -- H versus JOHNSON V -- 2MASS J. The composite sdB binaries from the sample presented in this article are shown in red triangles, and as comparison the sdB+WD binaries of \citet{Kupfer2015} are plotted in blue dots. The two systems that fall outside the main group of composite sdB binaries plotted in red squares are: PG\,1647+056 (open square) a sdB+K type binary and MCT\,2340$-$2806 (filled square) a sdB+G/F type binary. Both are spectroscopically confirmed.}
    \label{fig:color_color_diagram}
\end{figure}

The systems included in this sample have a magnitude range of V = 10 - 16 mag.
In Fig\,\ref{fig:color_color_diagram} a V-J vs J-H colour-colour diagram of the composite sdB sample is plotted, together with the sdB+WD systems from the short period sdB binary sample of \citet{Kupfer2015}. Only systems with reliable photometric measurements (error lower than 0.1 mag) are included in the plot. 
Both samples are clearly separated in the colour-colour diagram, with the binaries containing a MS component showing significant IR excess compared to the binaries with a WD companion. This division is known \citep[e.g.][]{Green2008}, and was used to select suitable targets from the GALEX and EC survey. The two composite sdB systems that do not fall within the main body of composite sdB binaries in Fig\,\ref{fig:color_color_diagram} are PG\,1647+056, a spectroscopically confirmed sdB+K binary \citep{Maxted2001} and MCT\,2340$-$2806, a spectroscopically confirmed sdB+G/F binary \citep{Lamontagne2000}.

\subsection{Misclassified targets}
Based on the UVES observations it was found that some targets from earlier IR excess based surveys have been misclassified as composite sdB binaries, they are briefly discussed below.

\begin{figure*}
    \includegraphics{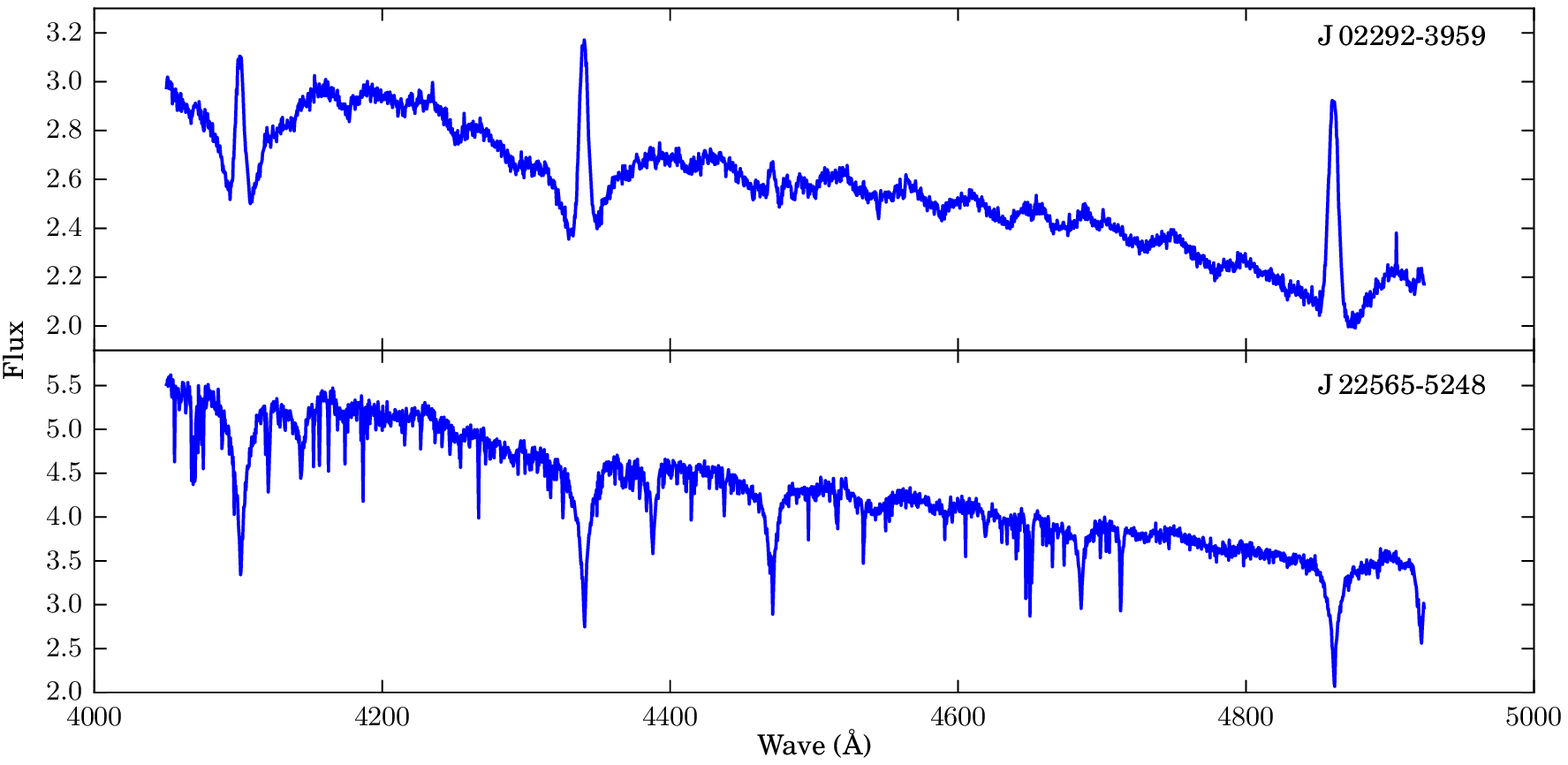}
    \caption{The BLUE arm UVES spectra of the cataclysmic variable J\,02292-3959 (top panel), and the He-sdOB J\,22565-5248 (bottom panel).}
    \label{fig:rejected_systems}
\end{figure*}

\begin{description}
\item[J02292-3959] (= GSC 07552-00389) is a recently discovered cataclysmic variable star of the VY Scl class \citep{Hummerich2014}, that shows dramatic and deep (3 to 4 magnitudes) fading events that can last for several years related to variations in the mass-transfer rate. Three spectra were obtained over a time span of two months, but all are remarkably similar. Broad, single peaked hydrogen and weak helium emission lines are visible in all our spectra originating from the accretion disk surrounding the white dwarf. However, blueward of the H$\beta$ line the white dwarf dominates the spectrum (the wings of the Balmer absorption lines are clearly visible), implying that the disk is relatively faint and so our spectra may well have been obtained during a fading event, when the mass transfer rate is decreased and the disk shrinks and fades. A BLUE arm UVES spectrum of this system is shown in the top panel of Fig\,\ref{fig:rejected_systems}.
\item[HE\,0505-2806] is a single sdB which shows no variations in radial velocity within the error bars. However, the spectrum is contaminated by light from a foreground wide binary containing at least one MS F or G type star. This contamination caused the misclassification as composite binary by \citet[identified as CD-28\,1974]{Vennes2011}. On aquisition images taken by EFOSC spectrograph at the NTT telescope in La Silla, the MS star is clearly visible, and is at a distance of 1.35'' from the sdB star. These images were taken at significantly better seeing than our UVES observations, which is why in our observations they apeared unresolved.
\item[EC\,12473-3046] classified by \citet{Copperwheat2011} as an sdB+MS binary is likely an sdB binary with an invisible companion. The sdB shows clear radial velocity variations with a maximum amplitude difference of $\bigtriangleup\,{\rm K} = 27 \pm 3$ \kms\ in  36 days. The cool companion however seems stationary within the error on the observations over a period of 2 years. Due to the poor seeing during the UVES observations, it is impossible to distinguish between the sdB binary and the cool star. It is possible that the cool star is bound to the inner binary on a very long period orbit, but in this case it would not have affected the evolution of the sdB. The inner binary must be an sdB with an unseen companion; either a white dwarf or an M-dwarf star, making the system a hierarchical triple system. EC\,12473-3046 would be the third candidate for a hierarchical triple system after PG\,1253+284 \citep{Heber2002} and J09510+03475 \citep{Kupfer2015}.
\item[PG\,1629+081] is an sdOB binary with an unseen companion. The amplitude of the RV variations indicates a short period binary.
\item[J07104+2333 and MCT\,2356-2655] are sdO binaries, and as such not of interest in this article.
\item[J22565-5248] is a single He-sdOB which does not show any radial velocity variation. A BLUE arm UVES spectrum of this system is shown in the bottom panel of Fig\,\ref{fig:rejected_systems}.
\item[PB 6148] is not subdwarf star but an F-type star, see also Sect.\,\ref{s:pb6148}.
\item[Balloon 110607002] was classified as an sdB+K system by \citep{Bixler1991}, but does not show any evidence of a subdwarf component in the UVES spectra, nor is there a strong UV flux detected by GALEX.
\end{description}

\begin{table*}
 \centering
   \caption{Extract from the VizieR catalog containing our sample of spectroscopically confirmed composite sdB binaries. The name is the simbad resolvable name or GALEX number. The reference is given in standard 19 digit format recognisable by ADS. In the VizieR catalogue 7 photometric measurements are given: GALEX FUV and NUV, Johnson B and V from APASS and 2MASS J, H and K. Errors on the photometric measurements are denoted by `e\_'} \label{tb:composite_sdbs}
   \begin{tabular}{llllllllll}
    \hline
    Name & RA & Dec & Class & Reference & FUV & e\_FUV & ... & Ks & e\_Ks \\ \hline
MCT 0000-1637  &  00:03:24.37 & $-$16:21:06.34   &  sdB+F/G   &  2000AJ....119..241L   &  12.687  &  0.007  &  &  12.596  & 0.026  \\
TON S 155      &  00:23:59.46 & $-$23:09:49.60   &  sdB+F7    &  2005A\&A...430..223L  &  14.750  &  0.019  &  &  15.426  & /      \\
MCT 0029-4054  &  00:31:36.27 & $-$40:37:54.61   &  sdB+F/G   &  2000AJ....119..241L   &  14.024  &  0.009  &  &  13.736  & 0.046  \\
PB 8555        &  01:10:12.64 & $-$14:07:56.98   &  sdB+F/G   &  2000AJ....119..241L   &  13.165  &  0.008  &  &  11.188  & 0.030  \\
PG 0110+262    &  01:13:14.89 & $+$26:27:30.99   &  sdB+G0V   &  2012ApJ...758...58B   &  12.767  &  0.004  &  &  12.215  & 0.022  \\
PB 6355        &  01:16:27.19 & $+$06:03:15.23   &  sdB+F     &  2012MNRAS.427.2180N   &  13.292  &  0.003  &  &  12.078  & 0.027  \\
EO Ceti        &  01:23:43.22 & $-$05:05:45.28   &  sdO/BV+F4V &  2012ApJ...758...58B   &  12.351  &  0.005  &  &  11.662  & 0.019  \\
HS 0127+3146   &  01:29:52.60 & $+$32:02:09.39   &  sdOB+MS   &  2003A\&A...400..939E  &  13.059  &  0.002  &  &  14.253  & 0.060  \\
PHL 1079       &  01:38:26.97 & $+$03:39:37.61   &  sdB+G7V   &  2012ApJ...758...58B   &  12.835  &  0.004  &  &  12.187  & 0.030  \\
    ...\\
    \hline \hline
   \end{tabular}

\end{table*}

\section{Observations}\label{s:observations}
Based on our composite sdB sample, explained in Sect.\,\ref{s:composite_sample}, we initiated a monitoring campaign using the UVES spectrograph at the VLT Kueyen telescope (UT2) on Cerro Paranal, Chile. UVES is a two-arm cross-dispersed echelle spectrograph. Which we used in standard dichroic-2 437+760 mode covering a wavelength range of 373 - 499 nm in the BLUE arm and 565 - 946 nm in the red arm. Using a slit width of 1 arcsec a resolution of about 40\,000 was reached. The targets are selected to cover all parts of the sky visible from Paranal, and are brighter than 14th magnitude. As the orbital period of these systems can be as long as 1300 days \citep{Vos2013}, observations in six consecutive semesters are necessary to cover the entire orbit. Currently observations during three semesters: P88 (1 Oct 2011 -- 31 Mar 2012), P93 (1 Apr 2014 -- 30 Sep 2014) and P96 (1 Oct 2015 -- 31 Mar 2016) have been obtained, thus no systems have complete orbital coverage yet. However, for 10 systems enough spectra have been obtained to create a master spectrum suitable for spectroscopic analysis of the cool companion and the sdB. These ten systems together with the number of observations and the semesters in which they are observed are given in Table\,\ref{tb:observations}. All UVES spectra were reduced using the UVES pipeline and the {\sc reflex} workflow engine \citep{reflex2013}.

\begin{table}
   \centering
   \caption{The number of obtained UVES spectra for each of the 10 targets analysed in this article and the semesters in which each target has been observed. Dates of the semesters are P88: 1\,Oct\,2011 -- 31\,Mar\,2012, P93: 1\,Apr\,2014 -- 30\,Sep\,2014 and P96: 1\,Oct\,2015 -- 31\,Mar\,2016} \label{tb:observations}
   \begin{tabular}{lcccc}
    \hline
    Object	& 	\multicolumn{4}{c}{Nr of Observations}   \\ 
    	& 	P88	&	P93	&	P96	&	Total	\\\hline \hline
    BPS CS 22890-74   &      0      &      3      &       0      &       3       \\
    HE 0430-2457      &      4      &      3      &       2      &       9       \\
    HE0505-2806       &      0      &      3      &       0      &       3       \\
    J02286-3625       &      0      &      3      &       3      &       6       \\
    J03154-5934       &      0      &      3      &       3      &       6       \\
    J03582-3609       &      0      &      3      &       3      &       6       \\
    JL 277            &      2      &      1      &       3      &       6       \\
    MCT 0146-2651     &      0      &      3      &       1      &       4       \\
    PB 6148           &      0      &      2      &       1      &       3       \\
    PG 1514+034       &      0      &      4      &       0      &       4       \\\hline
   \end{tabular}
\end{table}

To check the methods used to derive the spectral parameters of the cool companions of the UVES systems, three well studied long period sdB+MS binaries: Feige\,87, BD$+$34$^{\rm o}$1543 and BD$+$29$^{\rm o}$3070, are included in the article. These three systems, that we will refer to as the test systems, have been analysed in detail by \citet{Vos2013}. The test systems have been observed with the HERMES spectrograph attached to the 1.2m Mercator telescope on La Palma (ES). Furthermore there are low resolution (R$\approx$550) flux calibrated spectra of all three systems obtained with the Boller and Chivens (B\&C) spectrograph attached to the University of Arizona's 2.3\,m Bok telescope located on Kitt Peak. For BD$+$34$^{\rm o}$1543 there is also a low resolution (R$\approx$1800) spectrum available from LAMOST DR1 \citep{Luo2016}. These low resolution spectra were analysed with the {\sc XTgrid} code.

\section{Methods}\label{s:methods}
Spectral parameters for both the sdB and the cool companion are derived using two different approaches. The Grid Search in Stellar Parameters (GSSP) software package is used to determine the spectral parameters of the cool companion, using a master spectrum shifted to the rest frame of the companion. {\sc XTgrid} is used to derive spectral parameters for both the sdB and the cool companion using for each target the spectrum with the highest signal to noise ratio (SNR). In the following subsections both methods are described and their accuracy and precision is discussed. Furthermore three test systems are analysed using GSSP and {\sc XTgrid}, and the results are compared with literature values.

GSSP derives spectral parameters for the cool companion from a master spectrum without removing the contribution of the sdB companion. This latter contribution is handled by introducing a dilution parameter, and is in fact assumed non-variable over the wavelength range considered in the GSSP analysis. This approach might seem counter-intuitive compared to using a spectral model to subtract the sdB contribution as is used in, for example, the Versatile Wavelength Analysis (VWA) method. However, here we will show that the approach used in GSSP is valid over short spectral ranges where the sdB contribution is approximately constant.

\subsection{Master spectrum creation}\label{s:master_spectrum_creation}
To create the master spectrum for the cool companion, all spectra are shifted to the rest frame of the cool companion, after which they are summed up, and normalised. To determine the radial velocities of the cool companion a cross-correlation with a template spectrum was performed. The cross correlation is preformed in velocity space, using the red arm spectrum on the spectral regions 5940 - 6270 \AA\ and 6320 - 6530 \AA. The template is a synthetic spectrum calculated with the SynthV LTE-based radiative transfer code \citep{Tsymbal1996} based on estimated parameters. When the GSSP parameters are determined a new template spectrum is calculated using the derived parameters, and the master spectrum creation and GSSP analysis is repeated. There can be a little smearing of the spectral lines due to the error on the radial velocity measurements, but as the errors on the RVs of the cool companion are on the order of 0.5 km s$^{-1}$, this is negligible. The normalisation is done by manually selecting continuum points in the spectrum and fitting a spline function through those points. The resulting master spectra have a S/N of around 100. 

\subsection{GSSP}
The Grid Search in Stellar Parameters (GSSP) software package \citep{Tkachenko2015} is based on a grid search in the fundamental atmospheric parameters and (optionally) individual chemical abundances of the star (or binary stellar components) in question. It uses the method of atmosphere models and spectrum synthesis, which assumes a comparison of the observations with each theoretical spectrum from the grid. For the calculation of synthetic spectra, we use the SynthV LTE-based radiative transfer code \citep{Tsymbal1996} and a grid of atmosphere models pre-computed with the {\sc LLmodels} code \citep{Shulyak2004}.

GSSP allows for optimisation of six stellar parameters at a time: effective temperature (T$_{\rm eff}$), surface gravity (log$g$), metallicity ([Fe/H]), microturbulent velocity (v$_{\rm micro}$), projected rotational velocity (v$_{\rm r} \sin{i}$) and the dilution of the star which is defined as d = F$_{\rm MS}$ / (F$_{\rm MS}$ + F$_{\rm sdB}$). The synthetic spectra can be computed in any number of wavelength ranges, and each considered spectral interval can be from a few \AA{}ngstr\"{o}m up to a few thousand \AA{}ngstr\"{o}m wide. The grid of theoretical spectra is built from all possible combinations of the above mentioned parameters. Each spectrum from the grid is compared to the observed spectrum of the star and the chi-square merit function is used to judge the goodness of fit. The code delivers the set of best fit parameters, the corresponding synthetic spectrum, and the chi-square values for each grid point.

For the systems discussed in this paper, GSSP is used to fit the wavelength ranges of 5910 - 6270 and 6330 - 6510 \AA\ in the RED arm of UVES. Because the sdB star is most strongly contributing in the UV and in the blue, these two ranges give the best compromise between a high signal to noise, and a high contribution of the cool companion. There are also no spectral lines of the sdB component visible in these wavelength ranges. 

The parameter ranges of the model grid used in GSSP are given in Table\,\ref{tb:gssp_model_ranges}. In the GSSP fit, the micro turbulent velocity is kept fixed at 2.0 as the signal to noise of the spectra is not high enough to accurately determine this parameter. The projected rotational velocity is unconstrained and uses steps of 1 km s$^{-1}$. To determine the parameters of each system, first a coarse grid spanning the full range with steps of $\vartriangle$T$_{\rm eff}$ = 500 K,  $\vartriangle$log\,$g$ = 0.5 dex, $\vartriangle[$Fe/H$]$ = 0.4 dex and $\vartriangle$dilution = 0.1 is used. In the second run, a smaller ranged grid is calculated with the smallest available step size for each parameter.

\begin{table}
   \centering
   \caption{The parameter ranges of the model grid used in the GSSP fit together with the step size. The micro turbulent velocity is kept fixed, and v$_{\rm r} \sin{i}$ is the projected rotational velocity.} \label{tb:gssp_model_ranges}
   \begin{tabular}{lll}
    \hline
    Parameter   &   Range   &    Step \\ \hline \hline
    T$_{\rm eff}$ (K)      &   3000 -- 9000   &   100  \\
    log $g$ (dex)          &   2.5 -- 4.5     &   0.1  \\
    $[$Fe/H$]$ (dex)        &   -0.8 -- 0.8    &   0.1  \\
    v$_{\rm micro}$ (\kms) &   2.0            &   /    \\
    v$_{\rm r} \sin{i}$ (\kms)   &   1 -- 150       &   1    \\
    dilution               &   0.01 -- 1.00   &   0.03 \\\hline
   \end{tabular}
\end{table}

Together with the chi-squared value for each grid point, GSSP returns the chi-squared value corresponding with the 1 sigma range. This value is used to determine the 1 sigma error on each parameter. As this chi-square value is determined based on the complete grid, all possible parameter correlations are included in the final error. The important parameter correlations in the case of our stars are discussed in Section\,\ref{s:gssp_correlations}. As a rule of thumb, an estimate of the three sigma error is determined as $e_{3\sigma} = 2.2 \cdot e_{1\sigma}$. \citet{Tkachenko2015} reports that this error estimate corresponds to results obtained by Markov chain Monte Carlo simulations. The errors reported in this article are the $1\sigma$ and $3\sigma$ errors as described here.

\begin{figure*}
    \includegraphics[height=22cm]{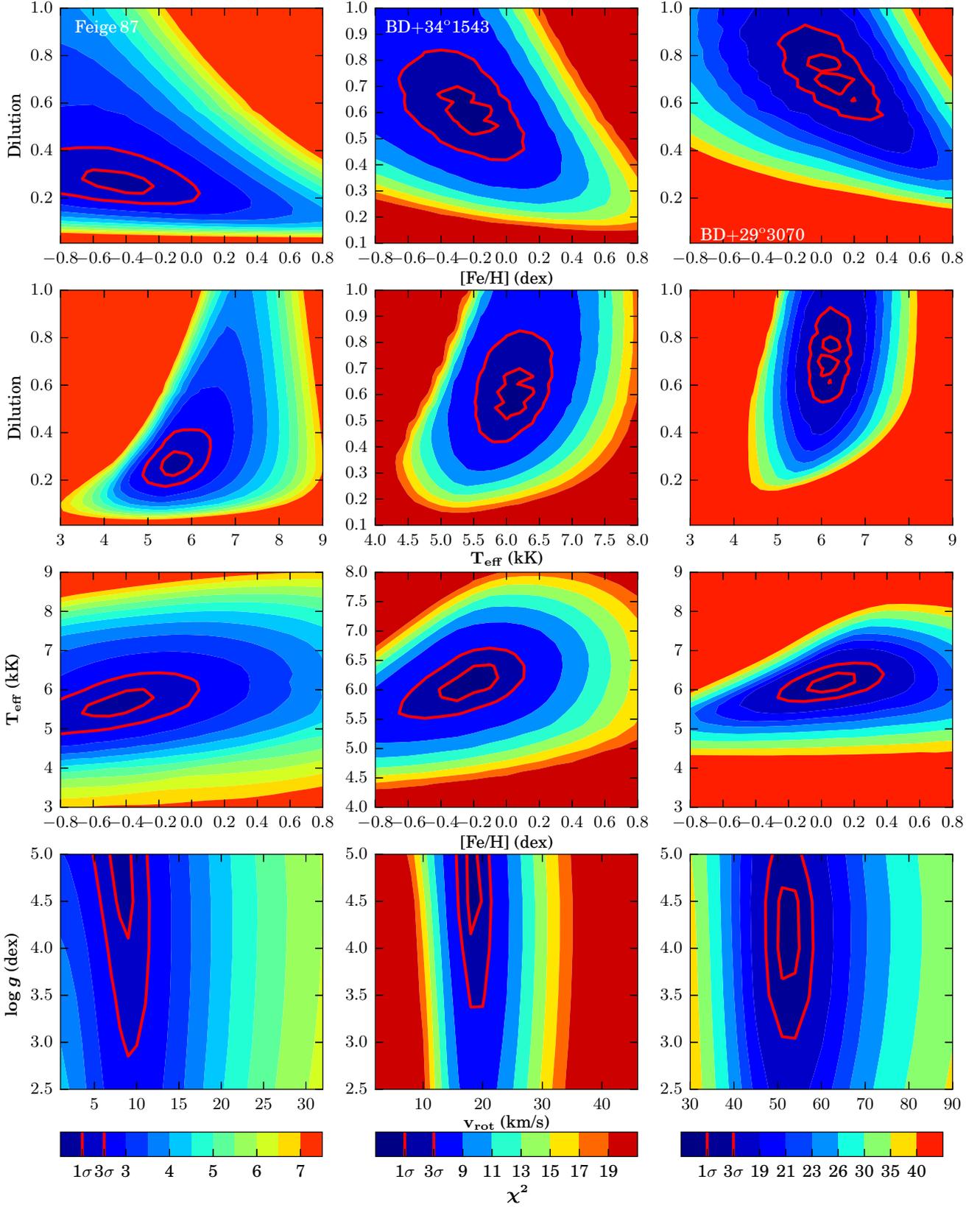}
    \caption{Chi-squared plots for different parameter combinations for three systems, left: Feige\,87 (slow rotator), middle: BD$+$34$^{\rm o}$1543 (medium rotator) and right: BD$+$29$^{\rm o}$3070 (fast rotator). The different colours represent different chi-squared values and the 1-sigma and 3-sigma confidence intervals are shown in red lines. See also section\,\ref{s:gssp_correlations}.}
    \label{fig:gssp_correlations}
\end{figure*}

\subsection{GSSP: accuracy and parameter correlations} \label{s:gssp_correlations}
To determine the accuracy of the GSSP code, and to get a better understanding of the different correlations between the five variables, a large detailed model grid spanning the entire parameter range given in Table\,\ref{tb:gssp_model_ranges} was calculated for the three test systems: BD$+$34$^{\rm o}$1543, BD$+$29$^{\rm o}$3070 and Feige\,87. These three sdB binaries with MS companions have been analysed in detail by \citet{Vos2013}, and are therefore ideal test cases for the GSSP software package. The chi-square maps of several parameter combinations are shown in Fig.\,\ref{fig:gssp_correlations}.

Based on these grids, it is clear that there is a strong negative correlation between metallicity and dilution. This correlation is expected as a decrease in the metallicity in the models would lead to weaker absorption lines, which can be countered by increasing the dilution factor. The effect of this correlation can be reduced by using enough different lines, which will respond differently on the changing metallicity, and can therefore not completely be compensated by changing the dilution.

In the slowly rotating Feige\,87 (V$_{\rm rot}$ = 8 $\pm$ 2 \kms), there is also a positive correlation between effective temperature and dilution. This correlation reduces greatly in strength with increasing V$_{\rm rot}$, and is completely absent in BD$+$29$^{\rm o}$3070 (V$_{\rm rot}$ = 52 $\pm$ 3 \kms). The faster rotating systems do show a positive correlation between T$_{\rm eff}$ and [Fe/H], which is almost invisible for Feige\,87. This correlation between T$_{\rm eff}$ and [Fe/H] would in combination with the correlation between [Fe/H] and dilution cancel out the correlation between T$_{\rm eff}$ and dilution. This explains why this latter correlation is absent in the faster rotating systems. 

Chi-square maps of the test systems for the parameter pairs with the strongest correlations are shown in the first three lines of Fig\,\ref{fig:gssp_correlations}. On this figure, the $1\sigma$ and $3\sigma$ confidence intervals are shown in thick red lines. The last line of this figure shows the chi-square map for $\log{g}$ versus v$_{\rm rot}$.

From the $1\sigma$ and $3\sigma$ contours shown in Fig\,\ref{fig:gssp_correlations} it is clear that most parameters can be determined accurately using the GSSP package. The only exception is the surface gravity. For all three test systems, the $3\sigma$ range for the surface gravity is about $\log{g} = [3.0 - 5.0]$. As the spectral lines in these types of stars have little dependency on the surface gravity, this is expected. 
For the systems observed with UVES, which have a lower signal to noise, the error on the $\log{g}$ will be even larger. The {\sc XTgrid} code can provide stronger constraints on the surface gravity, as it fits both components simultaneously, and the total flux of the sdB and cool companion must match the observations. Furthermore, when GAIA distances will be available, the spectral energy distribution can be used to constrain the radii of both components, and if the mass ratio is known, the surface gravities can be obtained with high precision.

\subsection{GSSP: precision}\label{s:gssp_comparison}

\begin{table*}
   \centering
   \caption{Comparison of spectroscopic parameters for three wide sdB binaries: BD$+$34$^{\rm o}$1543, BD$+$29$^{\rm o}$3070, Feige\,87 obtained using GSSP and VWA on high resolution spectra, SED fitting based on literature photometry and the {\sc XTgrid} code on low resolution spectra. The VWA, SED and {\sc XTgrid} results are taken from \citet{Vos2013}. The dilution is defined as d = F$_{\rm MS}$ / (F$_{\rm MS}$ + F$_{\rm sdB}$). See also section \ref{s:gssp_comparison}.} \label{tb:gssp_comparison}
   \begin{tabular}{llcccccc}
    \hline
    & Method  &       T$_{\rm eff}$   &       $\log{g}$       &       [Fe/H]          &       V$_{\rm micro}$ &  V$_{\rm rot} \sin{i}$  &       Dilution        \\
    &         &       (K)             &       (dex)           &       (dex)           &       (km s$^{-1}$)   &  (km s$^{-1}$)          &                       \\\hline\hline
\multirow{4}{*}{BD$+$34$^{\rm o}$1543} 
    & {\sc VWA    } &    6150 $\pm$ 150    &    4.20 $\pm$ 0.30   &    -0.24 $\pm$ 0.12  &    1.45 $\pm$ 0.25 &    17 $\pm$ 4     &    /               \\
    & {\sc SED    } &    6210 $\pm$ 220    &    4.20 $\pm$ 0.20   &    /                 &    /               &    /              &    0.67 $\pm$ 0.13 \\
    & {\sc XTgrid } &    6100 $\pm$ 300    &    4.11 $\pm$ 0.30   &    -0.40 $\pm$ 0.25  &    /               &    /              &    0.68 $\pm$ 0.10 \\
    & {\sc GSSP   } &    6100 $\pm$ 150    &    4.50 $\pm$ 0.50   &    -0.31 $\pm$ 0.10  &    2.0             &    19 $\pm$ 1     &    0.60 $\pm$ 0.05   \\\hline
\multirow{4}{*}{BD$+$29$^{\rm o}$3070} 
    & {\sc VWA    } &    6100 $\pm$ 200    &    4.30 $\pm$ 0.50   &    0.09 $\pm$ 0.21   &    1.50 $\pm$ 0.35 &    52 $\pm$ 4     &    /               \\
    & {\sc SED    } &    6570 $\pm$ 650    &    4.40 $\pm$ 0.40   &    /                 &    /               &    /              &    0.72 $\pm$ 0.12 \\
    & {\sc XTgrid } &    6026 $\pm$ 300    &    4.26 $\pm$ 0.30   &    0.08 $\pm$ 0.25   &    /               &    /              &    0.69 $\pm$ 0.10 \\
    & {\sc GSSP   } &    6170 $\pm$ 200    &    4.30 $\pm$ 0.40   &    0.05 $\pm$ 0.10   &    2.0             &    52 $\pm$ 2     &    0.70 $\pm$ 0.04 \\\hline
\multirow{4}{*}{Feige\,87}
    & {\sc VWA    } &    6175 $\pm$ 150    &    4.50 $\pm$ 0.60   &    -0.50 $\pm$ 0.25  &    1.15 $\pm$ 0.25 &     8 $\pm$ 3     &    /               \\
    & {\sc SED    } &    5840 $\pm$ 550    &    4.40 $\pm$ 0.25   &    /                 &    /               &     /             &    0.24 $\pm$ 0.12 \\
    & {\sc XTgrid } &    5675 $\pm$ 250    &    4.23 $\pm$ 0.35   &    -0.60 $\pm$ 0.20  &    /               &     /             &    0.37 $\pm$ 0.10 \\
    & {\sc GSSP   } &    5500 $\pm$ 230    &    4.50 $\pm$ 0.50   &    -0.50 $\pm$ 0.15  &    2.0             &     8 $\pm$ 1     &    0.30 $\pm$ 0.05 \\\hline
   \end{tabular}
\end{table*}

To check the results of the GSSP code for our systems we used it to analyse three wide sdB+MS binaries, BD$+$34$^{\rm o}$1543, BD$+$29$^{\rm o}$3070 and Feige\,87. These systems were observed with the HERMES spectrograph attached to the Mercator telescope, and the complete analysis was published by \citet{Vos2013}. The spectral parameters of the cool companions have been determined in three independent ways. Using the Versatile Wavelength Analysis code \citep[VWA,][]{Bruntt2002}, fitting synthetic binary spectra to the photometric spectral energy distribution (SED) and by fitting low resolution spectra using the {\sc XTgrid} code \citep{Nemeth2012}.

The VWA package fits abundances in a semi-automatic way. It first selects the least blended lines in the spectra, and determines the abundances from these lines by calculating synthetic spectra for each line while iteratively changing the input abundance until the equivalent widths (EW) of the observed and synthetic spectrum match. The input parameters (T$_{\rm eff}$, $\log{g}$ and v$_{\rm micro}$) are then manually changed until there is no correlation anymore between iron abundance and EW and excitation potential, and the \ion{Fe}{I} and \ion{Fe}{II} abundances match. The rotational velocity is determined by eye. The downside of this method is that the fit can reach a local minimum, and that the method is quite laborious when many stars need to be analysed.  VWA also doesn't allow for the fitting of a dilution factor, and would thus require several assumptions on the spectral parameters of the sdB star to disentangle the spectra. A detailed description of the VWA package can be found in \citet{Bruntt2004, Bruntt2008, Bruntt2009, Bruntt2010}. As VWA does not support the analysis of composite system, the HERMES spectra were first disentangled before a master spectrum was created which was then used as input for VWA. A detailed description of the analysis method is given in \citet{Vos2013}. 

The second set of parameters is derived using the SED fitting method. This method fits synthetic SEDs integrated from model atmospheres to photometric SEDs to determine the effective temperature and surface gravity of both components in the binary. The metallicity of the model atmospheres is solar, and v$_{\rm micro}$ and v$_{\rm r} \sin{i}$ can not be determined in this way. The dilution factor given in Table\,\ref{tb:gssp_comparison} is the average dilution in the wavelength range 5910 - 6510 \AA. A detailed description of this method is given in \citet{Degroote2011} and \citet{Vos2012, Vos2013}.

The third set of parameters is derived using {\sc XTgrid} on flux calibrated spectra obtained using the Boller and Chivens spectrograph attached to the University of Arizona's 2.3\,m Bok telescope located on Kitt Peak. These spectra have a 9 \AA\ resolution. {\sc XTgrid} fits both the sdB and cool companion in the system. v$_{\rm micro}$ and v$_{\rm r} \sin{i}$ cannot be determined using the low resolution data. A detailed explanation of the {\sc XTgrid} fit is given in Section\,\ref{s:xtgrid}.

For the GSSP analysis the exact same method as for the UVES systems was followed. For each of the three systems a master spectrum is created by shifting all HERMES spectra to the zero point of the cool companion. The same wavelength ranges of 5910 - 6270 and 6330 - 6510 \AA\ are then normalised by hand and used as input for GSSP. 

The comparison of the spectral parameters determined by these four methods are given in Table\,\ref{tb:gssp_comparison}. For almost all of the parameters, the GSSP results are in excellent agreement with the results of the other three methods. The only exception is the effective temperature determined by VWA of Feige\,87. As the GSSP results do match well with the results from the SED fit and the {\sc XTgrid} fit, it is likely that the effective temperature determined by VWA is wrong. This could have happened due to the VWA fit being stuck in a local minimum. 

This comparison of the GSSP results with the parameters determined using three other independent methods shows that the GSSP code can derive reliable spectral parameters for the cool companion stars in hot subdwarf binaries.

\subsection{{\sc XTgrid}}\label{s:xtgrid}
The preceeding sections have demonstrated that F and G type MS companions of sdB stars can be easily resolved from optical spectra as they have a similar flux contribution and a rich, distinct spectrum. This allows one to disentangle the composite spectra of such binaries from a single observation without knowing the radii or fluxes of the components. Such a binary decomposition is implemented in the steepest-descent chi-squared minimizing spectral fitting algorithm {\sc XTgrid}, which was first used to estimate the atmospheric parameters of the components in 29 composite spectra binaries from low resolution data  by \cite{Nemeth2012}. {\sc XTgrid} employs the 
NLTE model atmosphere code {\sc Tlusty} \citep{Hubeny1995} for the hot subdwarf component and interpolated observed templates or theoretical synthetic spectra  for the MS companion. Such a decomposition task is relatively easy with low-resolution, high S/N flux calibrated spectra. The binary spectrum can be fitted with the linear combination of the spectra of the two stellar components. In the case of echelle spectroscopy, such as the UVES sample presented here, an absolute flux calibration or continuum normalization is difficult to achieve. Therefore we applied an improved version of {\sc XTgrid} to the observations for the highest S/N spectrum of each binary system.

{\sc XTgrid} applies a global optimization in the full parameter space. It fits the $T_{\rm eff}$, $\log{g}$, the individual abundances of selected elements and the projected rotation velocity for the sdB star. For the companion, it fits the $T_{\rm eff}$, $\log{g}$, metallicity, alpha enhancement and projected rotation velocity. The two models are connected with a wavelength dependent dilution factor to render composite spectra from the individual stellar components. A radial velocity correction is included in the procedure and the component velocities are determined iteratively along with the surface parameters. The procedure does not check the luminosity ratio of the components, it seeks for the best possible reproduction of the observations by means of a global chi-squared minimization. In fact, in the case of HE\,0505-2806 we found an inconsistent luminosity ratio assuming the same distance to the components. This suggested that the system is not a physical binary and the companion is a foreground star, confirming the earlier results found by \cite{Nemeth2012}. 

Our models for the hot subdwarf star included opacities of H, He, C, N, O, Si and Fe in both the NLTE model atmosphere and the synthetic spectrum calculation. Mg was included only in the synthetic spectrum calculation in LTE conditions. The atmospheres were sampled in 50 depth points and converged to a relative change of 0.1\%.

We used the synthetic spectral library calculated with the {\sc Phoenix} model atmosphere code (version 1, \citealt{Husser2013}) for the cool companions. The {\sc Phoenix} grid covers both the entire parameter range we are interested in and the spectral range of UVES at high resolution including the Ca {\sc ii} near-infrared triplet. The grid also includes a microturbulent velocity which is parametrized with the stellar properties. We created a sub-grid from scaled solar metallicity models (no alpha enhancement) and limited the grid to $T_{\rm eff}=4000-8000$ K, $\log{g} = 3-5$ dex (cgs) and ${\rm [Fe/H]} > -2$ dex. In order to derive the binary parameters and cope with parameter correlations we found it necessary to fit the entire UVES range simultaneously. The BLUE arm constrains the hot subdwarf properties, while the REDL and REDU 
arms constrain the companion atmospheric parameters as well as the radial velocities and the dilution factor. Such a global approach is also preferable to cope with strong parameter correlations, that exist among metallicity, dilution and rotation. 

{\sc Xtgrid} starts out from an input model for the binary and improves the fit by successive approximations. To avoid getting trapped in local minima {\sc XTgrid} increases the parameter step sizes (the size of the simplex) regularly and re-starts its descent. The error calculation assumes that the steepest descent procedure reduces the correlations for the subdwarf and binary parameters to a minimum and therefore they can be treated separately. According to this the derived error bars are one-dimensional errors based on the $\chi^2$ statistics and neglecting parameter correlations. For the cool companions we used the $\chi^2$ distribution obtained during the last decomposition iteration to derive error bars. 

Although the iterative decomposition method employed by {\sc XTgrid} is more laborious and time consuming than the GSSP approach, a strong motivation to pursue in this direction was that {\sc XTgrid} can provide the atmospheric parameters for both components as well as their entire optical spectrum. This allows an improvement of the radial velocity determination for the hot subdwarf and refines the binary parameters further. 

\section{Atmospheric parameters}\label{s:results}
In Table\,\ref{tb:companion_parameters}, the spectral parameters derived using GSSP for the companions are given, together with the upper and lower one-sigma error values. For most systems only a lower error on the $\log{g}$ could be given as the fit reached the maximum value. Based on these results we have a broad range of main sequence companions with spectral types from K4V to F5V. As expected companions with a higher effective temperature contribute stronger to the composite spectra, which can be seen in the higher dilution factors. The rotational velocity ranges from a slowly rotating 6 km s$^{-1}$, to a very fast rotator with $v_{\rm r} \sin{i}$ = 90 km s$^{-1}$. For this small sample, there does not seem to be any correlation between the rotational velocity and the spectral type. The cool companions show a spread in metallicities varying from the very metal poor MCT\,0146-2651 with [Fe/H] = $-$0.80, to two systems, BPS\,CS\,22890--74 and J\,03582--3609, with a super solar metallicity.

The sdB components have a complicated formation history including diffusion processes that can cause chemical peculiarities \citep{Heber1998}. This makes it difficult if not impossible, to estimate the age of the sdBs from atmospheric parameters. When there has been little to no mass transfer in the history of the binary, the metallicity of the cool companion can be used as an age indicator. The spread in metallicity that we observe here suggest a wide range of ages for these systems.

\begin{table*}
   \centering
   \caption{Spectroscopic parameters for nine cool companions of the sdB binaries observed with UVES. The parameters determined using the GSSP code are given on the first line, while those determined by {\sc XTgrid} are given on the second line. The errors provided for GSSP are the $1\sigma$ errors, those for {\sc XTgrid} are the 60\% confidence intervals. The {\sc XTgrid} errors do not take parameter correlations into account, while the GSSP errors do. The dilution is the average dilution of the cool companion, d = F$_{\rm MS}$ / (F$_{\rm MS}$ + F$_{\rm sdB}$), in the wavelength range 5900-6500 \AA. (see also Sect.\,\ref{s:results})} \label{tb:companion_parameters}
   \begin{tabular}{llllll}
    \hline
    Name  &       T$_{\rm eff}$   &       $\log{g}$    &   [Fe/H]          &       V$_{\rm rot}$ sini   &       Dilution        \\
          &       (K)             &       (dex)        &   (dex)           &       (km s$^{-1}$)        &                       \\\hline\hline
  BPS\,CS\,22890--74   & $5630\ ^{+170}_{-220}$& $4.50\ ^{+0.00}_{-0.20}$& $0.10\ ^{+0.20}_{-0.10}$& $7\ ^{+1}_{-1}$& $0.45\ ^{+0.05}_{-0.05}$ \\ \noalign{\smallskip}
& $5650\ ^{+70}_{-70}$& $4.72\ ^{+0.23}_{-0.26}$& $0.50\ ^{+0.10}_{-0.10}$& $9\ ^{+0.3}_{-0.3}$& $0.42\ ^{+0.01}_{-0.01}$ \\ \noalign{\smallskip}\noalign{\smallskip}
      HE\,0430--2457   & $4620\ ^{+170}_{-200}$& $4.50\ ^{+0.00}_{-0.50}$& $-0.35\ ^{+0.30}_{-0.25}$& $30\ ^{+3}_{-3}$& $0.20\ ^{+0.05}_{-0.05}$ \\ \noalign{\smallskip}
& $4790\ ^{+250}_{-250}$& $5.00\ ^{+0.40}_{-0.20}$& $-0.50\ ^{+0.10}_{-0.30}$& $36\ ^{+5}_{-5}$& $0.24\ ^{+0.01}_{-0.01}$ \\ \noalign{\smallskip}\noalign{\smallskip}
      J\,02286--3625   & $5330\ ^{+180}_{-180}$& $4.50\ ^{+0.00}_{-0.50}$& $-0.15\ ^{+0.10}_{-0.15}$& $90\ ^{+5}_{-3}$& $0.30\ ^{+0.05}_{-0.05}$ \\ \noalign{\smallskip}
& $5090\ ^{+150}_{-90}$& $3.80\ ^{+0.20}_{-0.20}$& $-0.32\ ^{+0.10}_{-0.10}$& $99\ ^{+8}_{-10}$& $0.40\ ^{+0.01}_{-0.01}$ \\ \noalign{\smallskip}\noalign{\smallskip}
      J\,03154--5934   & $6080\ ^{+190}_{-190}$& $4.40\ ^{+0.10}_{-0.35}$& $-0.30\ ^{+0.15}_{-0.10}$& $23\ ^{+1}_{-1}$& $0.55\ ^{+0.05}_{-0.05}$ \\ \noalign{\smallskip}
& $6070\ ^{+150}_{-100}$& $4.34\ ^{+0.10}_{-0.10}$& $-0.03\ ^{+0.20}_{-0.10}$& $24\ ^{+0.3}_{-0.5}$& $0.55\ ^{+0.02}_{-0.02}$ \\ \noalign{\smallskip}\noalign{\smallskip}
      J\,03582--3609   & $6030\ ^{+220}_{-220}$& $4.20\ ^{+0.30}_{-0.40}$& $0.00\ ^{+0.15}_{-0.15}$& $30\ ^{+2}_{-2}$& $0.75\ ^{+0.10}_{-0.10}$ \\ \noalign{\smallskip}
& $6200\ ^{+110}_{-240}$& $4.50\ ^{+0.10}_{-0.07}$& $0.50\ ^{+0.30}_{-0.30}$& $31\ ^{+1.2}_{-1.2}$& $0.76\ ^{+0.01}_{-0.01}$ \\ \noalign{\smallskip}\noalign{\smallskip}
             JL\,277   & $6550\ ^{+200}_{-200}$& $4.50\ ^{+0.00}_{-0.50}$& $-0.40\ ^{+0.15}_{-0.15}$& $33\ ^{+2}_{-2}$& $0.60\ ^{+0.05}_{-0.10}$ \\ \noalign{\smallskip}
& $6270\ ^{+80}_{-130}$& $4.11\ ^{+0.15}_{-0.15}$& $-0.50\ ^{+0.10}_{-0.10}$& $32\ ^{+2.3}_{-1.3}$& $0.62\ ^{+0.01}_{-0.01}$ \\ \noalign{\smallskip}\noalign{\smallskip}
     MCT\,0146--2651   & $5910\ ^{+160}_{-220}$& $4.50\ ^{+0.00}_{-0.15}$& $-0.80\ ^{+0.00}_{-0.10}$& $6\ ^{+2}_{-2}$& $0.40\ ^{+0.05}_{-0.05}$ \\ \noalign{\smallskip}
& $5800\ ^{+100}_{-100}$& $4.50\ ^{+0.20}_{-0.30}$& $-1.00\ ^{+0.10}_{-0.10}$& $5.6\ ^{+0.8}_{-0.6}$& $0.42\ ^{+0.01}_{-0.01}$ \\ \noalign{\smallskip}\noalign{\smallskip}
            PB\,6148   & $6540\ ^{+160}_{-120}$& $4.50\ ^{+0.00}_{-0.50}$& $-0.35\ ^{+0.05}_{-0.10}$& $15\ ^{+1}_{-1}$& $0.94\ ^{+0.03}_{-0.03}$ \\ \noalign{\smallskip}
& $6250\ ^{+50}_{-50}$& $3.73\ ^{+0.07}_{-0.10}$& $-0.46\ ^{+0.10}_{-0.10}$& $15\ ^{+0.4}_{-0.4}$& $1.00\ ^{+0.01}_{-0.00}$ \\ \noalign{\smallskip}\noalign{\smallskip}
      PG\,1514$+$034   & $5630\ ^{+270}_{-350}$& $4.50\ ^{+0.00}_{-0.50}$& $-0.50\ ^{+0.30}_{-0.15}$& $18\ ^{+2}_{-2}$& $0.44\ ^{+0.10}_{-0.10}$ \\ \noalign{\smallskip}
& $5900\ ^{+130}_{-185}$& $5.00\ ^{+0.25}_{-0.10}$& $0.00\ ^{+0.10}_{-0.10}$& $20\ ^{+0.7}_{-0.8}$& $0.32\ ^{+0.01}_{-0.01}$ \\ \noalign{\smallskip}\noalign{\smallskip}
\hline
   \end{tabular}
\end{table*}

\begin{table}
   \centering
   \caption{T$_{\rm eff}$ and $\log{g}$ of the sdB components derived using the {\sc XTgrid} code. The upper and lower errors are the 60\% confidence intervals. PB\,6148 is not included in this table as it does not contain a hot subdwarf. (see Sect.\,\ref{s:results})} \label{tb:xtgrid_sdB_results}
   \begin{tabular}{lll}
    \hline
    Name  &    T$_{\rm eff}$ sdB   &    $\log{g}$ sdB   \\
          &    (K)             &    (dex)        \\\hline\hline
  BPS\,CS\,22890--74   & $22890 ^{+1320}_{-320}$& $5.24 ^{+0.18}_{-0.07}$ \\ \noalign{\smallskip}
       HE\,0430-2457   & $26500 ^{+500}_{-600}$& $5.38 ^{+0.10}_{-0.10}$ \\ \noalign{\smallskip}
      J\,02286--3625   & $28790 ^{+430}_{-440}$& $5.38 ^{+0.08}_{-0.08}$ \\ \noalign{\smallskip}
      J\,03154--5934   & $26740 ^{+1000}_{-780}$& $5.39 ^{+0.13}_{-0.14}$ \\ \noalign{\smallskip}
      J\,03582--3609   & $31330 ^{+2530}_{-1130}$& $5.80 ^{+0.21}_{-0.60}$ \\ \noalign{\smallskip}
             JL\,277   & $25410 ^{+320}_{-940}$& $5.44 ^{+0.04}_{-0.23}$ \\ \noalign{\smallskip}
     MCT\,0146--2651   & $36460 ^{+530}_{-610}$& $5.72 ^{+0.07}_{-0.08}$ \\ \noalign{\smallskip}
        PG\,1514+034   & $34590 ^{+750}_{-800}$& $5.53 ^{+0.14}_{-0.13}$ \\ \noalign{\smallskip}
\hline
   \end{tabular}
\end{table}

\begin{table*}
   \centering
   \caption{The surface abundances for the sdB components determined with {\sc XTgrid}. The errors are the 60\% confidence intervals. PB\,6148 is not included in this table as it does not contain a hot subdwarf. (see Sect.\,\ref{s:results})} \label{tb:xtgrid_abundances}
   \begin{tabular}{llllllll}
    \hline
    Name  &    \multicolumn{6}{c}{$\log(nX/nH) \pm \log(1 + \delta nX/nX)$}       \\
          &    He   &    C   &    N   &    O   &    Mg   &    Si  &  Fe  \\\hline\hline \noalign{\smallskip}
  BPS\,CS\,22890--74   & $-3.07 ^{+0.29}_{-0.36}$&  $< -4.34 $              &  $< -3.70 $              &  $< -3.65 $              &  $< -4.62 $              &  $< -5.22 $              & $-4.54 ^{+0.75}_{-2.93}$ \\ \noalign{\smallskip}
       HE\,0430-2457   & $-2.59 ^{+0.19}_{-0.12}$& $-4.06 ^{+0.20}_{-0.45}$& $-4.38 ^{+0.36}_{-0.80}$& $-4.31 ^{+0.30}_{-0.68}$&  $< -4.25 $              & $-4.52 ^{+0.48}_{-0.48}$&  $< -4.50 $               \\ \noalign{\smallskip}
      J\,02286--3625   & $-2.55 ^{+0.18}_{-0.16}$& $-4.70 ^{+0.54}_{-0.68}$& $-4.44 ^{+0.27}_{-0.31}$& $-4.53 ^{+0.16}_{-0.34}$&  $< -5.18 $              &  $< -6.25 $              &  $< -4.77 $               \\ \noalign{\smallskip}
      J\,03154--5934   & $-2.80 ^{+0.41}_{-0.23}$&  $< -3.96 $              & $-4.64 ^{+0.54}_{-1.00}$& $-3.81 ^{+0.18}_{-0.61}$&  $< -4.64 $              &  $< -5.66 $              &  $< -4.31 $               \\ \noalign{\smallskip}
      J\,03582--3609   & $-1.23 ^{+0.40}_{-0.67}$& $-3.24 ^{+0.72}_{-1.66}$&  $< -3.09 $              & $-3.02 ^{+0.26}_{-0.59}$&  $< -2.02 $              & $-3.86 ^{+0.46}_{-1.04}$&  $< -3.47 $               \\ \noalign{\smallskip}
             JL\,277   & $-2.23 ^{+0.09}_{-0.20}$& $-4.59 ^{+0.40}_{-0.72}$& $-4.47 ^{+0.27}_{-0.61}$& $-3.65 ^{+0.15}_{-0.42}$&  $< -4.97 $              &  $< -5.34 $              &  $< -4.27 $               \\ \noalign{\smallskip}
     MCT\,0146--2651   & $-1.57 ^{+0.13}_{-0.10}$&  $< -5.24 $              &  $< -4.37 $              &  $< -4.65 $              &  $< -4.34 $              &  $< -5.38 $              & $-4.36 ^{+0.53}_{-0.49}$ \\ \noalign{\smallskip}
        PG\,1514+034   & $-1.90 ^{+0.18}_{-0.17}$&  $< -4.79 $              & $-4.74 ^{+0.99}_{-1.11}$&  $< -4.61 $              &  $< -4.40 $              &  $< -5.41 $              & $-5.24 ^{+1.54}_{-2.66}$ \\ \noalign{\smallskip}
\hline
   \end{tabular}
\end{table*}

\begin{figure*}
    \includegraphics{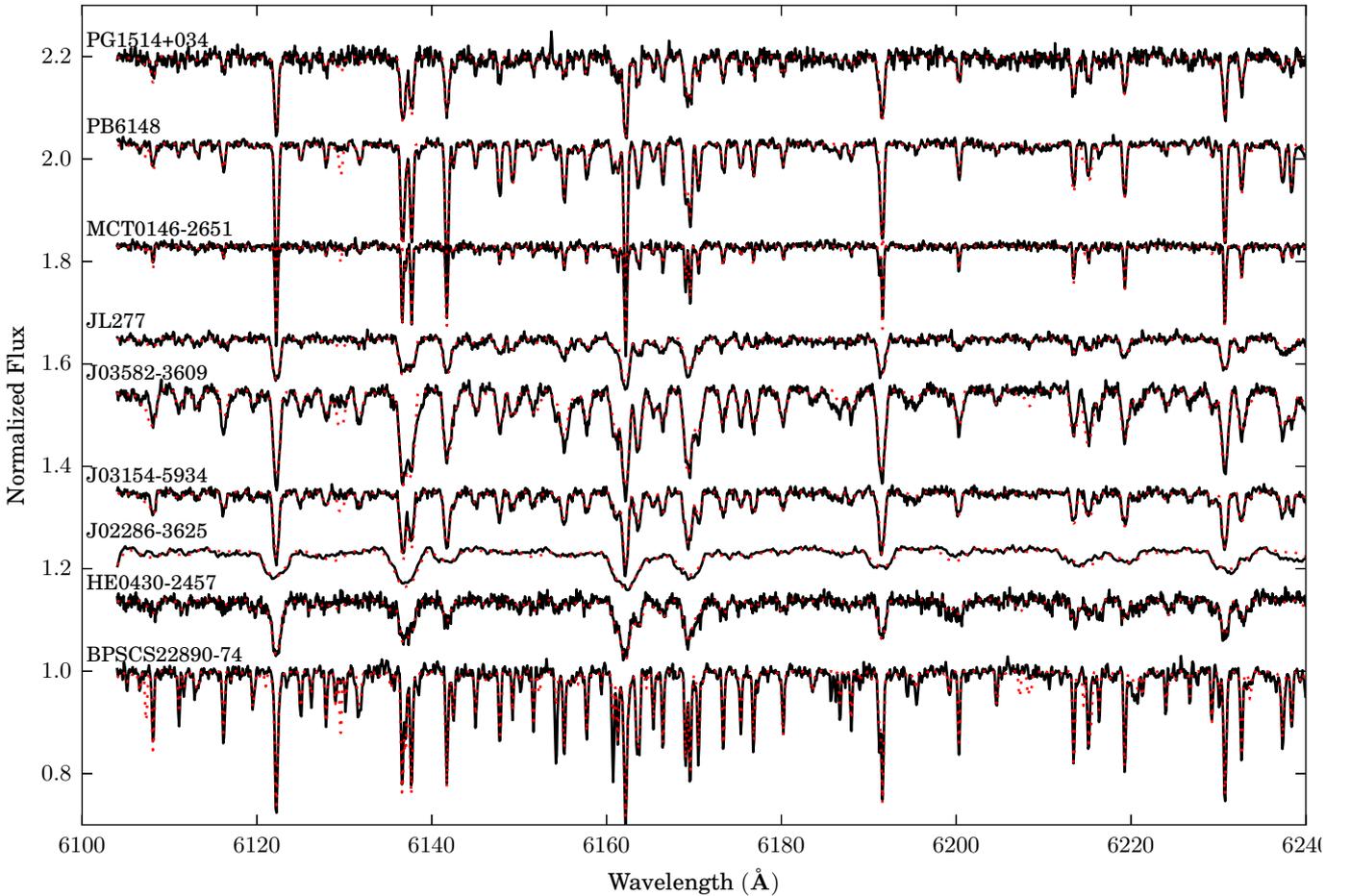}
    \caption{The observed normalized spectra (black full line) and best fitting GSSP model (red dotted line) for a section of the wavelength range used to determine the spectroscopic parameters with GSSP.}
    \label{fig:gssp_results}
\end{figure*}

\begin{figure*}
    \includegraphics{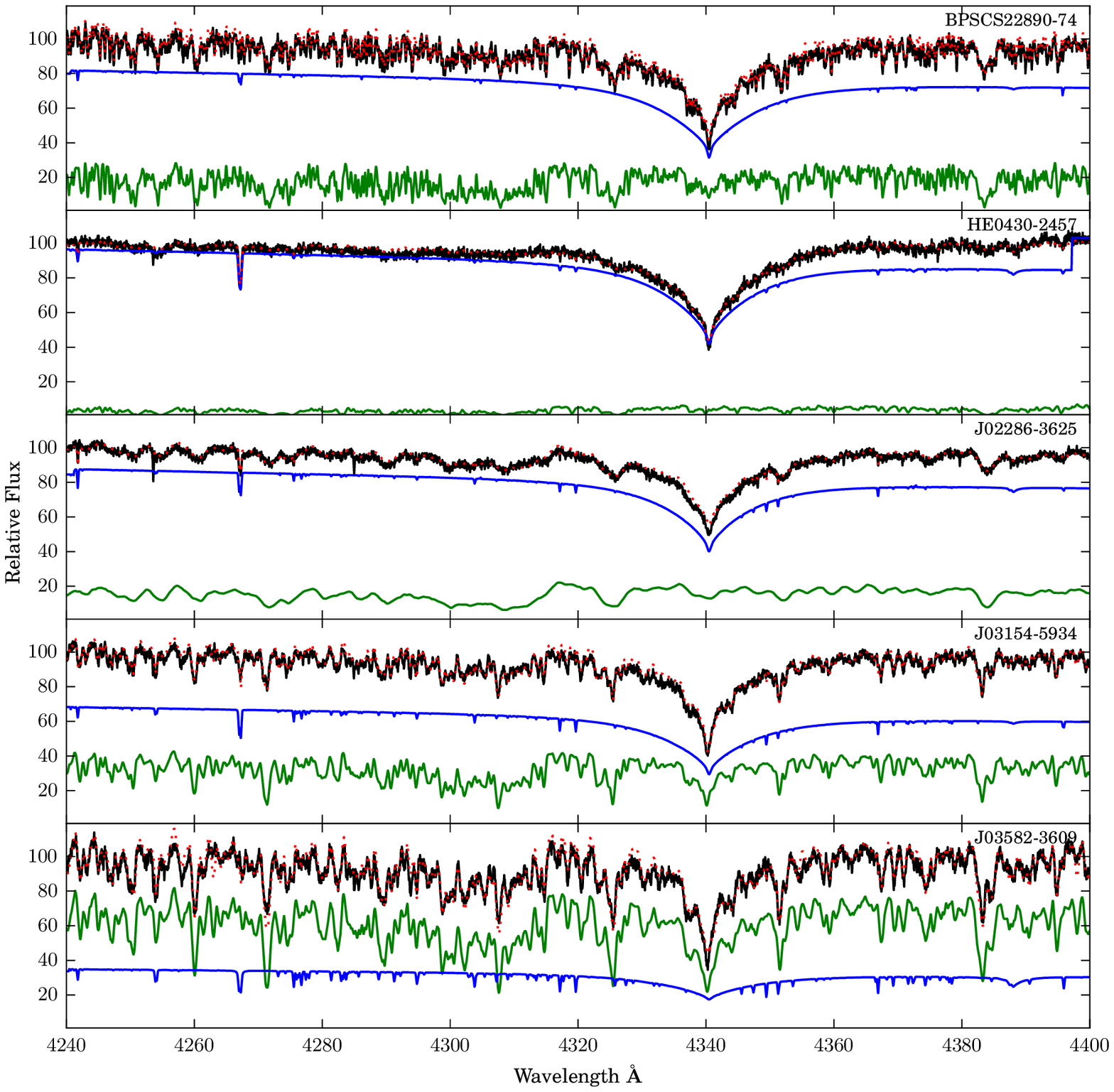}
    \caption{The observed UVES spectra (black full line) and best fitting {\sc XTgrid} model. The cool companion contribution is shown in green line, the sdB contribution is blue line, and the binary model in red dotted line. The flux is scaled so that the average continuum flux of the observed spectrum equals 100 for easy comparison between the systems. Interactive figures are available at: \url{http://astroserver.org/KW32YZ/}}
    \label{fig:xtgrid_results_1}
\end{figure*}

\begin{figure*}
    \includegraphics{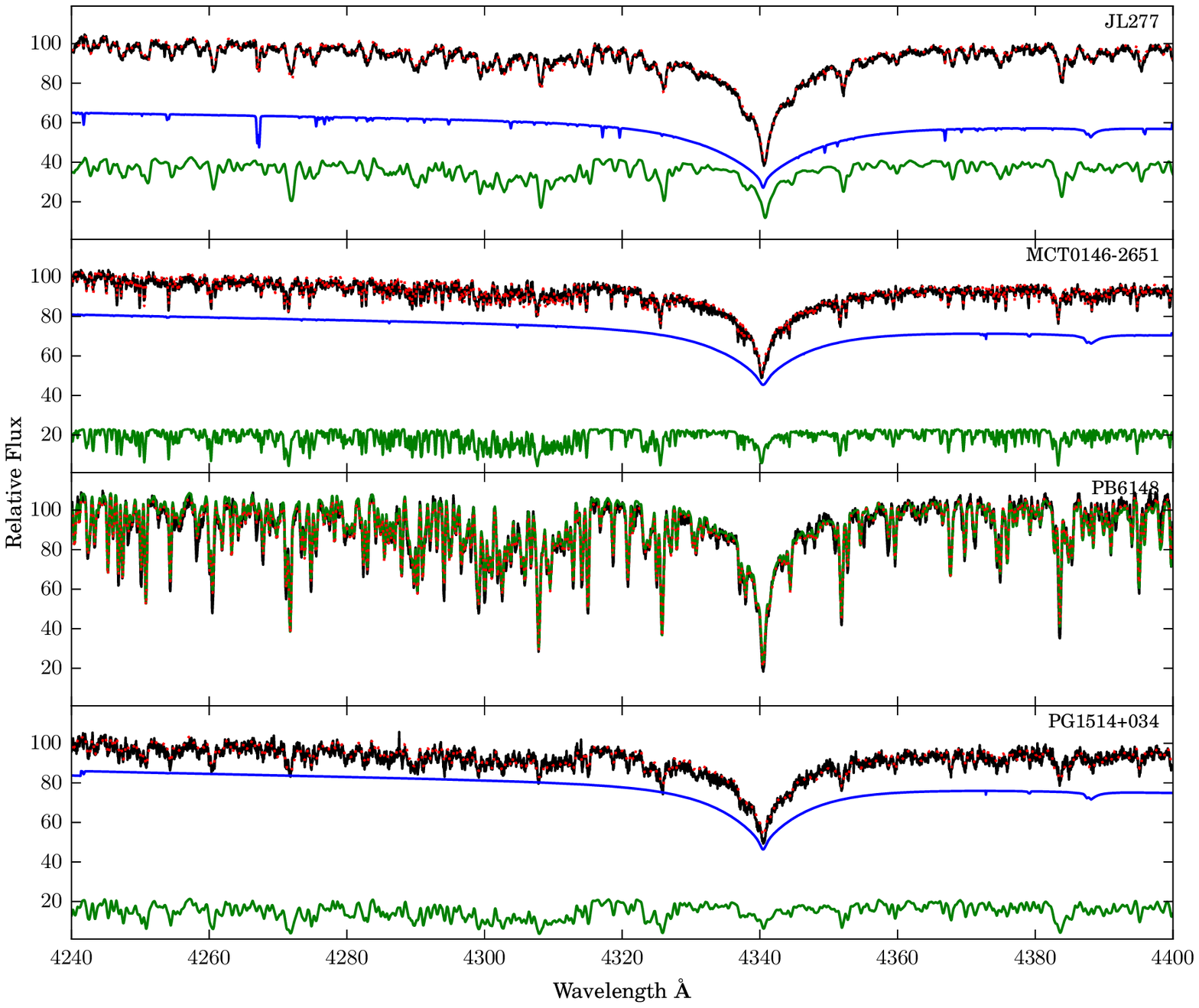}
    \caption{The observed UVES spectra (black full line) and best fitting {\sc XTgrid} model. The cool companion contribution is shown in green line, the sdB contribution is blue line, and the binary model in red dotted line. The flux is scaled so that the average continuum flux of the observed spectrum equals 100 for easy comparison between the systems. Interactive figures are available at: \url{http://astroserver.org/KW32YZ/}}
    \label{fig:xtgrid_results_2}
\end{figure*}

The {\sc XTgrid} results for the companion stars are given on the second line for each system in Table\,\ref{tb:companion_parameters}. The effective temperature and surface gravity of the sdB components determined using {\sc XTgrid} are given in Table\,\ref{tb:xtgrid_sdB_results}. The abundances of He, C, N, O, Mg, Si and Fe for the sdB component are given in Table\,\ref{tb:xtgrid_abundances}. PB\,6148 is missing in the latter two tables as it does not have an sdB component (see Sect.\,\ref{s:pb6148}). A section of the total fitted region with {\sc XTgrid} for each system is shown in Figures\,\ref{fig:xtgrid_results_1} and \ref{fig:xtgrid_results_2}. These figures show the best fitting binary spectra, as well as the contribution of the sdB and the cool companion.

When comparing the {\sc XTgrid} results with the GSSP results one has to take into account that the errors given for the {\sc XTgrid} parameters are not including the correlations between the parameters, and as such are smaller then the GSSP errors. For the nine systems discussed here, the atmospheric parameters determined with the two methods are in good agreement. The largest difference between the two methods is the derived metallicity. For about half of the systems, the {\sc XTgrid} results find a significantly higher metallicity. This difference is a result of the different model atmospheres employed by the two methods, and is discussed in the next section. 

\subsection{sdB He sequence}
{\sc XTgrid} was originally designed to fit single-lined hot subdwarf spectra and binary decomposition was added later to be able to fit the considerable number of hot subdwarfs in composite spectrum binaries. The high resolution of UVES and the adequate S/N ratio allows us to separate the components and derive precise atmospheric parameters for both the subdwarf and the cool companion. Such a task is usually not possible from low resolution data. 

\begin{figure}
    \includegraphics[width=3.464in]{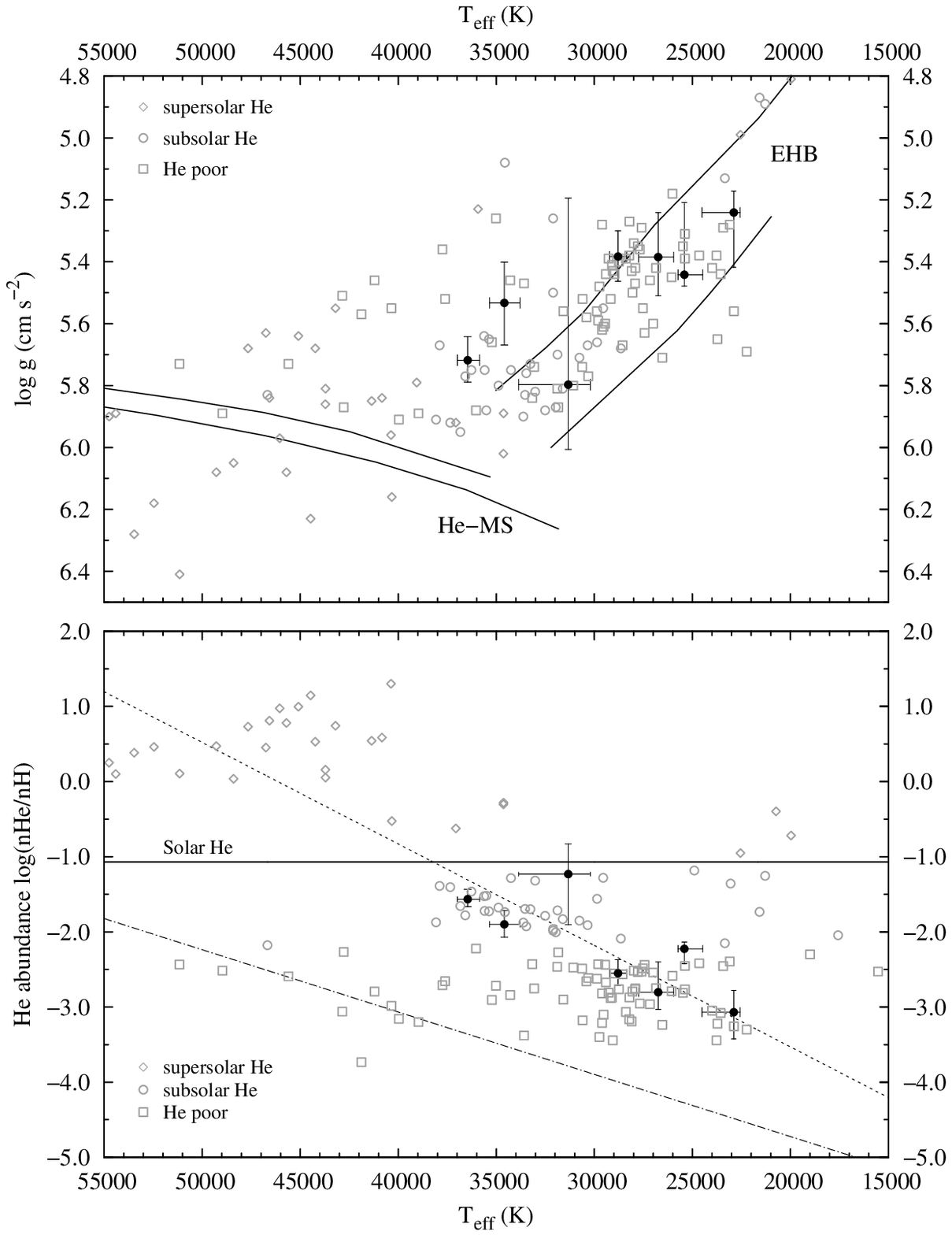}
    \caption{Hot subdwarfs from the UVES sample in the $T_{\rm eff}-\log{g}$ (upper panel) and $T_{\rm eff}-\log(n{\rm He}/n{\rm H})$ (bottom panel) planes. The gray data points provide a reference for the subdwarf atmospheric properties and are taken from the GALEX sample \citep{Nemeth2012}. The EHB and He main-sequence (He-MS) are marked in the top panel. The horizontal line in the bottom panel shows the solar helium abundance. The dashed line in the $T_{\rm eff}-\log(n{\rm He}/n{\rm H})$ panel is the correlation taken from an independent analysis by \citet{Edelmann2003}. The dash-dotted line (bottom panel) is the lower boundary of the surface helium abundances found in the GALEX sample.}
    \label{fig:sdb_teff_he_uves}
\end{figure}

\citet{Edelmann2003} found a positive correlation between $T_{\rm eff}$ and $\log(n{\rm He}/n{\rm H})$ of hot subdwarfs and that two such He sequences exist. We found that all subdwarfs in the UVES sample follow this trend and clearly belong to the He-rich sequence. Past investigations revealed that sdB stars along the He-rich sequence appear in two major clumps and these clumps correlate with the binary and pulsation properties of the stars. \citet{Nemeth2012} found in the GALEX sample that most hot subdwarfs in composite spectra belong to the higher temperature, gravity and helium abundance group. Fig.\,\ref{fig:sdb_teff_he_uves} shows that all sdB binaries observed with UVES can be associated with both of these clumps, however, the current sample is too small to draw any conclusion. Our efforts in collecting a larger sample will reveal the relation between the EHB morphology and composite spectrum binaries.

The surface gravity of the two sdOB systems (MCT\,0146--2651 and PG\,1514+034) suggests that these subdwarfs have started evolving off the Extreme Horizontal Branch (EHB) while their helium abundance is still consistent with EHB stars.

The spectral analysis provided reliable measurements for $T_{\rm eff}$, $\log{g}$ and for the He abundance as listed in Table\,\ref{tb:xtgrid_sdB_results} . 
The fits to binary systems have more free parameters, therefore, even though the lines of Mg and Si are clearly present in the spectra we can report only upper limits of their abundances in Table\,\ref{tb:xtgrid_abundances} at the given level of confidence. Despite the fact that metal abundances are difficult to measure from optical spectra, the abundances we found are comparable to literature values (\citealt{Otoole2006}, \citealt{Blanchette2008}). Metals are at sub-solar abundances in all stars. Carbon and oxygen show a large scatter from star to star, while nitrogen is the most abundant light metal. Iron is very close to the solar iron abundance in agreement with the results of \cite{Geier2013}.

\subsection{Metallicity difference between the {\sc LLmodels} and {\sc Phoenix} models}
Although the atmospheric parameters of the cool companion determined with GSSP and {\sc XTgrid} correspond well overall, there is a discrepancy in the metallicity for about half of the systems. In the case of BPS\,CS\,22890--74, HE\,0430--2457, J\,03154--5934, J\,03582--3609 and PG\,1514$+$034, the metallicity determined with {\sc XTgrid} is significantly higher than the GSSP metallicity. The difference beeing roughly 0.4 dex. The reason for this offset may be related to the different models used for the cool companion.

\begin{figure*}
    \includegraphics{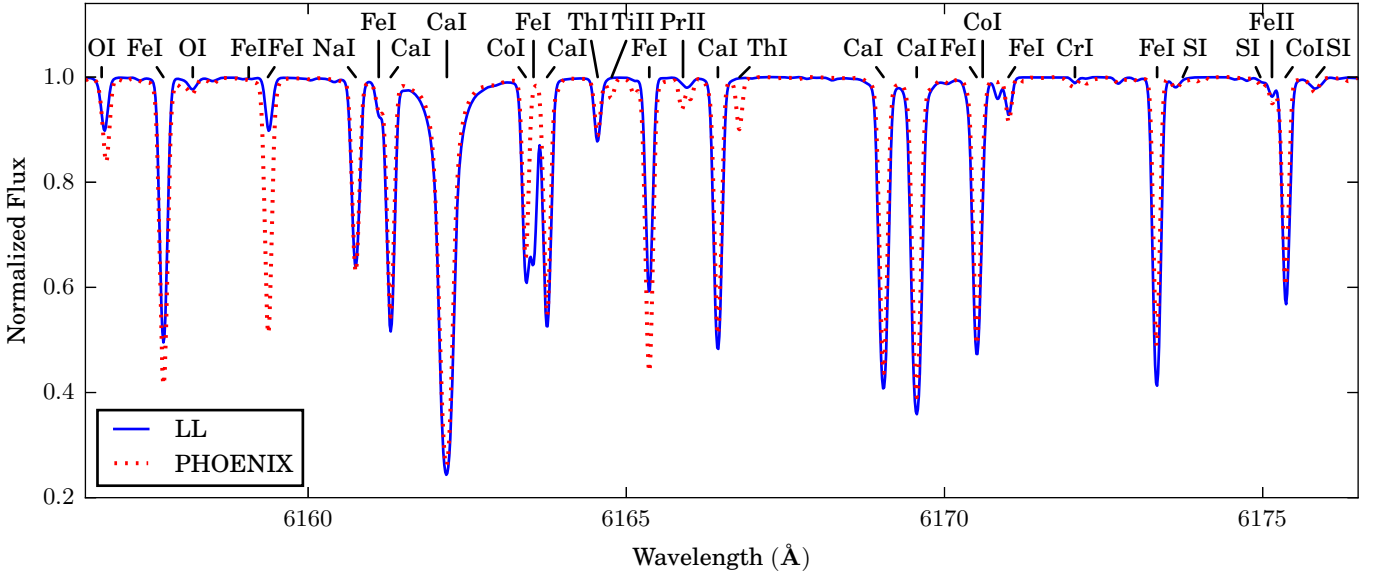}
    \caption{Comparison of a LL synthetic spectrum (blue full line) with a {\sc Phoenix} synthetic spectrum (red dotted line). Both spectra are caculated for T$_{\rm eff}$ = 5700 K, $\log{g}$ = 4.0 dex, [Fe/H] = 0.0, [$\alpha$/Fe] = 0.0 and convolved to fit the UVES resolution. The stronger lines are identified above the spectra. There is a clear difference in the strength of several Fe lines visibile, with some lines being cleary present in one model but totaly absent in the other.}
    \label{fig:LL_vs_PHOENIX}
\end{figure*}

GSSP uses Line by Line opacity ({\sc LLmodels}, \citealt{Shulyak2004}), which are local thermal equilibrium (LTE) plane-parallel stellar model atmospheres for early and intermediate type stars using the Vienna Atomic Line Database (VALD, \citealt{Ryabchikova2015}). In the current analysis we used {\sc XTGrid} with {\sc Phoenix} models \citep{Husser2013}, which also assume LTE but the modeling geometry, atomic data input and numeric algorithms are likely different than in {\sc LLmodels}. In Fig.\,\ref{fig:LL_vs_PHOENIX} a synthetic spectrum calculated both with the {\sc LLmodels} and {\sc Phoenix} models is shown. This spectrum has an T$_{\rm eff}$ = 5700 K, $\log{g}$ = 4.0 dex, [Fe/H] = 0.0, [$\alpha$/Fe] = 0.0 and is convolved to fit the resolution of our UVES spectra. In the wavelength range shown,  it is clear that there is a significant difference in some of the Fe lines between the two models, with some lines being cleary present in one model but totaly absent in the other. Other lines as for example \ion{Ca}{i} are exactly the same in both models. With this in mind we attribute the different metalicities reported by both fitting algorithms to the differences in the synthetic atmospheres they use.

Even though the effective temperature, surface gravity, rotation and dilution derived for the nine systems given in Table\,\ref{tb:companion_parameters} is comparable between the LL and {\sc Phoenix} models, the metallicity is not. This comparison shows that the atmosphere models used to determine atmospheric parameters have an influence on those parameters, and that if a comparison of the atmospheric parameters of different stars is made, it should ideally be done using the same atmosphere models.

This metallicity difference does not show up when comparing the {\sc XTgrid} results for the three test systems in Section\,\ref{s:gssp_comparison}. This is because those systems were analysed using an earlier version of {\sc XTgrid} with {\sc Atlas} models \citep{Bertone2004} in the range of 3700-7200 \AA, instead of the {\sc Phoenix} models used in the analysis of the UVES systems.

\subsection{PB\,6148}\label{s:pb6148}
PB\,6148 was identified by \citet{Berger1980} as an sdOB binary with an F type companion based on low resolution spectroscopy (dispersion $\sim$ 80 \AA/mm) with the ``D'' spectrograph at the 80\,cm telescope at Observatoire de Haute-Provence. However, in our high-resolution UVES observations, there is no spectral signature of an sdB component visible. Even though there is no indication of an sdB component in the spectra, the spectral fit of the F-type star with GSSP indicates that there is a component present contributing about 5\% of the total light, while the {\sc XTgrid} fit attributes 10\% of the total light to the unidentified companion. 

In total three UVES spectra of PB\,6148 were obtained on HJD 2455908, 2455934 and 2456851. For these spectra the radial velocities of the F-type star were determined using a cross correlation with a template spectra as explained in Section\,\ref{s:master_spectrum_creation}. The F-type star shows radial velocity variations with a maximum amplitude of 2.5 km/s. This variation is five times larger than the uncertainty on the radial velocity measurements, indicating that PB\,6148 is likely a binary system. There is also a very weak line visible at $\lambda$ 5875 \AA, which might be a \ion{He}{i} line. Radial velocity measurements of this line alone show a movement in antiphase with that of the F-type star, but the errors are of the same order as the amplitude. Based on the available data, we can not identify the secondary component of PB\,6148.

\section{Mass accretion by the companions}\label{s:accretion}
The evolutionary state of some companions to sdB stars has been used as an argument that the companions could not have accreted much mass during the RLOF phase. If the companion star is slightly evolved, it cannot have accreted much matter, as a star with a lower initial mass would not have had enough time to evolve \citep[see for example][]{Vos2012}. However, no attempt has been made to quantify the amount of accreted mass. 

By measuring the current rotational velocity and comparing this to the initial synchronised state, an estimate of the accreted mass can be made based on the amount of angular momentum that needs to be transferred onto the companion to reach the observed rotation velocity. The angular momentum of the companion based on the rotational velocity is given by:
\begin{equation}
 \Omega = r_{\rm g} M_{\rm c} R_{\rm c}^2 \cdot \omega,
\end{equation}
where $r_{\rm g}$ is the gyration radius of the companion, and $M_{\rm c}$ and $R_{\rm c}$ are the mass and radius of the companion. $\omega$ is the angular velocity, which can be calculated as $\omega = v_{\rm rot} / R = 2 \pi / P_{\rm spin}$, with $P_{\rm spin}$ the spin period. The change in angular momentum before and after mass transfer is then:
\begin{equation}
 \bigtriangleup \Omega = r_{\rm g} M_{\rm c} R_{\rm c}^2 \left( \frac{v_{\rm rot, f}}{R_{\rm c}} - \frac{2 \pi}{P_{\rm spin, i}} \right), \label{eq_delta_omega}
\end{equation}
where $P_{\rm spin, i}$ is the initial spin period of the companion before the onset of RLOF and $v_{\rm rot, f}$ is the observed rotational velocity. As the binary would have been circularised and synchronised before the onset of RLOF, the initial spin period equals the orbital period before RLOF $P_{\rm spin, i} = P_{\rm orb, i}$. The orbital period before RLOF is estimated between 500 and 900 days, as this corresponds to the separation necessary to initiate RLOF near the tip of the red giant branch \citep{Vos2015}

Eq. \ref{eq_delta_omega} assumes that the radius of the companion does not change significantly between the onset of RLOF, and the moment when the system is observed. The timescales involved here are short compared to the MS lifetime of the companion, thus this assumption is valid in the case where the amount of mass accreted by the companion is low. 

\subsection{Angular momentum transfer during RLOF}
The amount of angular momentum transfer differs depending on whether the mass flow directly hits the companion (ballistic mass transfer), or first forms an accretion disk and only then gets accreted onto the companion (Keplerian disk mass transfer). To determine which regime is applicable in this case we calculate the minimum distance of the mass stream to the companion following the fitting formula from \citet{Ulrich1976} based on \citet{Lubow1975}:
\begin{equation}
 R_{\rm min} = 0.0425\ a\ ( q + q^2)^{1/4},
\end{equation}
where $a$ is the separation and $q = M_{\rm c} / M_{\rm sdB}$. If the distance between the mass stream and the centre of mass of the companion is larger than its radius, an accretion disk will be formed. In the case of wide sdB binaries, we calculate the minimum distance to the companion for systems with $P_{\rm i}$ = 500 - 900 days, $P_{\rm f}$ = 700 - 1300 days, an sdB progenitor mass between 1.5 and 2.0 M$_{\odot}$, an sdB mass of 0.47 M$_{\odot}$, and a companion with a mass of 1 M$_{\odot}$. For these parameters, the minimum distance between the mass stream and the companion varies between 15 and 33 R$_{\odot}$. This is in all cases much larger than the radius of the companion, and the mass lost from the sdB progenitor will thus never directly hit the companion. 

The angular momentum accreted by the companion from an accretion disk is the angular momentum of the inner radius of the Keplerian accretion disk around the companion, which is taken at the surface of the companion:
\begin{equation}
 \delta\Omega_{\rm acc} = \sqrt{ G M_{\rm c} R_{\rm c} } \cdot \delta M_{\rm acc}, \label{eq_omega_accreted}
\end{equation}
where $\delta\Omega_{\rm acc}$ is the accreted angular momentum per unit mass accreted by the companion $\delta M_{\rm acc}$ and G is the gravitational constant. 

\subsection{Angular momentum loss during the sdB phase.}
To check if the current observed rotational velocity of the companion is the same as its rotational velocity after the end of RLOF we compare the sdB lifetime with the synchronisation timescales for the companion. Using the formalism of \citet{Zahn1977} the synchronisation timescale of the companion is given by:
\begin{equation}
 t_{\rm syn} = \frac{1}{6 q^2 k_2} \left( \frac{M_{\rm c} R_{\rm c}}{L_{\rm c}} \right)^{1/3} \frac{I_{\rm c}}{M_{\rm c} R_{\rm c}^2} \left(\frac{a}{R_{\rm c}}\right)^6,
\end{equation}
where $q$ = M$_{\rm sdB}$ / M$_{\rm c}$, k$_2$ is the apsidal motion constant, L$_{\rm c}$ is the luminosity of the companion, $I_{\rm c} = r_{\rm g} M_{\rm c} R_{\rm c}^2$ is the moment of inertia and $a$ is the separation. \citet{Brooker1955} calculated apsidal motion constants using polytrope models. For a main sequence star with polytropic index $n=3$, $k_2=0.0144$. Gyration radii were calculated by \citet{Claret1989}, also using polytrope models, which for main sequence stars result in $r_g = 0.076$.

Assuming a solar type companion, and an average separation of $a = 500 R_{\odot}$, the synchronisation timescale exceeds the Hubble time. 
Synchronisation will thus not affect the rotational velocity of the companions after the end of RLOF.

Synchronisation is not the only way to lose spin angular momentum. In this derivation we ignore other angular-momentum-loss effects, for example due to stellar winds or magnetic fields, as these effects are difficult to quantify.

\subsection{Accreted mass}\label{s:accreted_mass}
By linking the accreted angular momentum per unit mass (eq. \ref{eq_omega_accreted}), to the observed change in angular momentum (eq. \ref{eq_delta_omega}), the amount of mass accreted by the companion can be calculated as:
\begin{equation}
  M_{\rm acc} = r_{\rm g} \sqrt{\frac{M_{\rm c} R_{\rm c}^3}{G}} \cdot \left( \frac{v_{\rm rot,f}}{R_{\rm c}} - \frac{2 \pi}{P_{\rm orb,i}} \right).
\end{equation}
As only spectral parameters of the companions are known, we use the average masses and radii based on the spectral types to determine the accreted mass. The spectral types, masses and radii were matched to the observations using the results of \citet{Pecaut2013} and the empirical mass-radius relation of \citet{Demircan1991}. In deriving the accreted mass, an error of 10\% was assumed on the masses and radii.

The inclination of these systems is not known, therefore we calculated a lower limit on the accreted mass using $i = 90^{\rm o}$, and an upper limit using $i = 21^{\rm o}$, which is the lowest inclination measured for a wide sdB binary. In Table\,\ref{tb:mass_accretion_results}, the accreted mass in both cases together with the spectral type and derived masses and radii are given for all analysed systems.

From these results it is clear that there is very little mass accretion onto the companions. The fastest rotator, J\,02286-3625, would only need to accrete at most 0.04 $M_{\odot}$ in order to
reach its observed rotational velocity if this system is observed at an inclination of 21$^{\rm o}$. On average these systems have accreted between 0.005 and 0.013 $M_{\odot}$ respectively in the $i = 90^{\rm o}$ and $i = 21^{\rm o}$ case.

\begin{table*}
   \centering
   \caption{The amount of mass that needs to be accreted onto the companions to reach the observed rotational velocities for two possible inclinations. The masses and radii are based on the spectral type. The minimum duration of mass accretion $\tau_{\rm min}$ is given for M$_{\rm acc}$ ($i = 21^{\rm o}$). See Sect.\,\ref{s:accretion} for details.} \label{tb:mass_accretion_results}
   \begin{tabular}{lcccr@{ $\pm$ }lr@{ $\pm$ }lr@{ $\pm$ }lr@{ $\pm$ }l}
    \hline
    Name  &       Spectral   &       Mass      &   Radius        &  \multicolumn{2}{c}{v$_{\rm rot} \sin{i}$}  &       \multicolumn{2}{c}{M$_{\rm acc}$ ($i = 90^{\rm o}$)}  &       \multicolumn{2}{c}{M$_{\rm acc}$ ($i = 21^{\rm o}$)}   &      \multicolumn{2}{c}{$\tau_{\rm min}$}  \\
          &       Type       &   ($M_{\odot}$) &  ($R_{\odot}$)  &  \multicolumn{2}{c}{(km s$^{-1}$)}          &       \multicolumn{2}{c}{($M_{\odot}$)}                     &       \multicolumn{2}{c}{($M_{\odot}$)}                      &      \multicolumn{2}{c}{(yr)}              \\\hline\hline
    BPS\,CS\,22890-74    & G6V & 0.97 & 1.03  &  7   &  1  & 0.0012 & 0.0002 & 0.0034 & 0.0006 & 3.2  & 0.6 \\
    HE\,0430-2457        & K4V & 0.73 & 0.79  &  30  &  3  & 0.0040 & 0.0005 & 0.0111 & 0.0015 & 13.7 & 2.4 \\
    J\,02286-3625        & K0V & 0.87 & 0.93  &  90  &  5  & 0.0141 & 0.0016 & 0.0389 & 0.0043 & 40.9 & 6.3 \\
    J\,03154-5934        & F9V & 1.14 & 1.20  &  23  &  1  & 0.0047 & 0.0005 & 0.0129 & 0.0013 & 10.6 & 1.5 \\
    J\,03582-3609        & G0V & 1.08 & 1.14  &  30  &  2  & 0.0057 & 0.0007 & 0.0160 & 0.0019 & 13.8 & 2.2 \\
    JL\,277              & F5V & 1.33 & 1.39  &  33  &  2  & 0.0077 & 0.0009 & 0.0217 & 0.0026 & 15.2 & 2.4 \\
    MCT\,0146-2651       & G1V & 1.07 & 1.13  &  6   &  2  & 0.0011 & 0.0004 & 0.0032 & 0.0012 & 2.7  & 1.0 \\
    PG\,1514+034         & G6V & 0.97 & 1.03  &  18  &  2  & 0.0031 & 0.0004 & 0.0086 & 0.0013 & 8.2  & 1.4 \\
\hline
   \end{tabular}
\end{table*}

\subsection{Duration of accretion}
The maximum rate at which a star can accrete matter is defined by the Eddington luminosity. If one assumes that the total luminosity of the star originates from energy deposited onto the star by the accreted matter, the Eddington accretion limit is given by:
\begin{equation}
 \dot{M}_{\rm edd} = \frac{4 \pi c}{\kappa} R_{\rm c},
\end{equation}
where c is the speed of light and $\kappa$ is the opacity which in the simplest case can be taken from Thompson scattering $\kappa$ = 0.4 cm$^2$ g$^{-1}$. If applied on the accretion masses derived here (Sect.\,\ref{s:accreted_mass}), the minimum time necessary to accrete this matter is on the order of years to decades. The timescales calculated for the amount of mass accreted in the case the inclination is $i = 21^{\rm o}$, are given in Table\,\ref{tb:mass_accretion_results}. This is the minimum duration, and the real time is likely longer as the accretion takes place from an accretion disk while the Eddington limit is an upper limit derived for spherically symmetric accretion, and the Thompson scattering opacity is not the best approximation for a MS star. However, the timescales derived here allow the prediction of thermal timescale mass loss for red giants.

\subsection{Evolutionary considerations}
The determination of the mass accretion by the companion based on its rotational velocity is only valid if there has indeed been a mass-transfer phase in the history of the system. In the case of very wide initial orbits, no mass transfer would take place during the RGB phase, and the measuered rotational velocity of the cool companion would be its initial rotational velocity. The upper limit on the initial period to allow for mass transfer during the RGB phase is around 1000 days. Depending on the mass-loss formalism used, these systems can reach orbital periods of 1600 days or more \citep{Chen2013, Vos2015}. Thus, if the current orbital period is shorter than about 1600 days, it is likely that the systems was formed through stable RLOF.

There is no complete orbital coverage for the studied systems yet, but based on the radial velocity amplitudes, the estimated periods of the systems discussed here are compatible with those of the already solved systems (P $\approx 700 - 1300$ d, \citealt{Vos2017}). Therefore it is very likely that these systems went through a mass-loss phase. The only alternative formation channel would be the double WD merger scenario in a triple system consisting of a close double WD binary in a wide orbit with an MS star, but it is unlikely for this scenario to be exclusively observed.

As the majority of the mass lost by the sdB progenitor is not accreted by the companion, it has to leave the system. This supports the proposed existence of circumbinary disks in wide sdB binaries. Based on binary evolution models, the sdB progenitor will lose more than 0.5 M$_{\odot}$, part of which might be lost through the outer Lagrange point \citep{Vos2015}. This mass will form a CB disk in which dust formation can take place. Photo-evaporation will remove the gas from the disk when the sdB star ignites its He core, but larger dust particles could survive for the duration of the core He-burning phase.

\section{Conclusions}
In this article we have presented a sample of confirmed composite sdB binaries containing 148 systems. This sample is constructed based on a literature review extended with systems from the GALEX and EC surveys which were confirmed spectroscopically as composite sdB binaries. In many of these composite sdB binaries, the subdwarf component is formed through stable RLOF, resulting in a long period binary with an MS companion. As this sample was constructed to study the post-RLOF population, all known short-period sdB binaries were removed. Based on observations with the UVES spectrograph, 8 systems with an invalid composite sdB classification have been reclassified.  

Based on the composite sdB sample a long term observing campaign of wide sdB binaries was started using the UVES spectrograph at the VLT. No complete orbits have been obtained yet. However, for nine systems enough spectra have been collected to create a master spectrum with a sufficiently high S/N to perform a spectroscopic analysis of the cool companion and the sdB star.

The GSSP code used to derive the spectroscopic parameters of the cool companion was tested using three systems studied in the literature, BD$+$34$^{\rm o}$1543, BD$+$29$^{\rm o}$3070 and Feige\,87. The GSSP parameters have been compared with parameters derived by independent methods, confirming the accuracy and precision of the GSSP code. This comparison shows that even though the contribution of the sdB component to the spectrum in GSSP is treated as a wavelength independent dilution factor. If the wavelength range is short enough, the parameters derived with GSSP for the cool companion are reliable.

The derived parameters of the 9 cool companions in this article show a diversity in spectral types, varying from mid K to mid F, as well as a variety in metallicity from very metal poor to slightly super solar metallicity. A third interesting parameter is the rotational velocity of the companions, with slowly rotating stars at $v_{\rm r} \sin{i}$ $<$ 10 km s$^{-1}$ to a fast rotator with $v_{\rm r} \sin{i}$ = 90 km s$^{-1}$. 

The parameters for the cool companion derived with GSSP and with {\sc XTgrid} match well within their error bars. Only the metallicity is an exception. For several of the systems, the metallicity derived with {\sc XTgrid} is larger than the metallicity derived with GSSP. We consider that the most likely reason for this is the atmosphere models used by the two codes. This shows that when comparing metallicities of systems, it is important to know what atmosphere models were used to derive them. 

The rotational velocity of the companions was used to estimate the amount and duration of accretion during the RLOF phase. By comparing the original synchronised rotation before RLOF to the observed rotational velocity, the amount of mass accretion can be estimated. Assuming that there is no significant angular momentum loss after the end of RLOF, we find that all systems have accreted very little mass, varying from 10$^{-3}$ to 10$^{-2}$ M$_{\odot}$. This is in agreement with the advanced evolutionary state of the companions in some wide sdB binaries \citep{Vos2012}. Such an evolution state could not have been reached if the companion had accreted a lot of mass during RLOF, as its original mass would have been too low to evolve off the MS. By using the maximum accretion rate based on the Eddington luminosity, the minimum accretion duration is found to be short, on the order of years. This lower limit does allow for the predicted thermal timescale mass loss during RLOF from a red giant.

Nine systems analysed in this paper are part of a larger sample of 37 sdB binaries that is currently beeing observed using the UVES spectrograph. The two methods that will be used to analyse these composite sdB binaries, grid search in stellar parameters (GSSP) and {\sc XTgrid}, were described in detail. The orbital and spectral analysis of the full sample will provide a statistically viable sample to study the formation mechanisms of wide sdB binaries, and the necessary eccentricity pumping processes to explain the currently observed period - eccentricity distribution.

\section*{Acknowledgements}
JV acknowledges financial support from FONDECYT in the form of grant number 3160504.
This research has used the services of {\sc Astroserver.org} under reference KW32YZ.
Thanks to John Telting for observing J03397-2459 with the Nordic Optical Telescope.
This publication makes use of data products from the Two Micron All Sky Survey, which is a joint project of the University of Massachusetts and the Infrared Processing and Analysis Centre/California Institute of Technology, funded by the National Aeronautics and Space Administration and the National Science Foundation.
This research was made possible through the use of the AAVSO Photometric All-Sky Survey (APASS), funded by the Robert Martin Ayers Sciences Fund.
This research makes use of data products from the Galaxy Evolution Explorer (GALEX), which is a NASA Small Explorer (www.galex.caltech.edu). The mission was developed in cooperation with the Centre National d' \'{E}tudes Spatiales (CNES) of France and the Korean Ministry of Science and Technology.




\bibliographystyle{mnras}
\bibliography{bibliogaphy} 

\begin{thebibliography}{}
\makeatletter
\relax
\def\mn@urlcharsother{\let\do\@makeother \do\$\do\&\do\#\do\^\do\_\do\%\do\~}
\def\mn@doi{\begingroup\mn@urlcharsother \@ifnextchar [ {\mn@doi@}
  {\mn@doi@[]}}
\def\mn@doi@[#1]#2{\def\@tempa{#1}\ifx\@tempa\@empty \href
  {http://dx.doi.org/#2} {doi:#2}\else \href {http://dx.doi.org/#2} {#1}\fi
  \endgroup}
\def\mn@eprint#1#2{\mn@eprint@#1:#2::\@nil}
\def\mn@eprint@arXiv#1{\href {http://arxiv.org/abs/#1} {{\tt arXiv:#1}}}
\def\mn@eprint@dblp#1{\href {http://dblp.uni-trier.de/rec/bibtex/#1.xml}
  {dblp:#1}}
\def\mn@eprint@#1:#2:#3:#4\@nil{\def\@tempa {#1}\def\@tempb {#2}\def\@tempc
  {#3}\ifx \@tempc \@empty \let \@tempc \@tempb \let \@tempb \@tempa \fi \ifx
  \@tempb \@empty \def\@tempb {arXiv}\fi \@ifundefined
  {mn@eprint@\@tempb}{\@tempb:\@tempc}{\expandafter \expandafter \csname
  mn@eprint@\@tempb\endcsname \expandafter{\@tempc}}}

\bibitem[\protect\citeauthoryear{{Aznar Cuadrado} \& {Jeffery}}{{Aznar
  Cuadrado} \& {Jeffery}}{2001}]{Aznar2001}
{Aznar Cuadrado} R.,  {Jeffery} C.~S.,  2001, \mn@doi [\aap]
  {10.1051/0004-6361:20010068}, \href
  {http://cdsads.u-strasbg.fr/abs/2001A%26A...368..994A} {368, 994}

\bibitem[\protect\citeauthoryear{{Barlow}, {Wade}, {Liss}, {{\O}stensen}  \&
  {Van Winckel}}{{Barlow} et~al.}{2012}]{Barlow2012}
{Barlow} B.~N.,  {Wade} R.~A.,  {Liss} S.~E.,  {{\O}stensen} R.~H.,   {Van
  Winckel} H.,  2012, \mn@doi [\apj] {10.1088/0004-637X/758/1/58}, \href
  {http://adsabs.harvard.edu/abs/2012ApJ...758...58B} {758, 58}

\bibitem[\protect\citeauthoryear{{Berger} \& {Fringant}}{{Berger} \&
  {Fringant}}{1980}]{Berger1980}
{Berger} J.,  {Fringant} A.-M.,  1980, \aaps, \href
  {http://adsabs.harvard.edu/abs/1980A%26AS...39...39B} {39, 39}

\bibitem[\protect\citeauthoryear{{Bertone}, {Buzzoni}, {Ch{\'a}vez}  \&
  {Rodr{\'{\i}}guez-Merino}}{{Bertone} et~al.}{2004}]{Bertone2004}
{Bertone} E.,  {Buzzoni} A.,  {Ch{\'a}vez} M.,   {Rodr{\'{\i}}guez-Merino}
  L.~H.,  2004, \mn@doi [\aj] {10.1086/422486}, \href
  {http://adsabs.harvard.edu/abs/2004AJ....128..829B} {128, 829}

\bibitem[\protect\citeauthoryear{{Bixler}, {Bowyer}  \& {Laget}}{{Bixler}
  et~al.}{1991}]{Bixler1991}
{Bixler} J.~V.,  {Bowyer} S.,   {Laget} M.,  1991, \aap, \href
  {http://cdsads.u-strasbg.fr/abs/1991A%26A...250..370B} {250, 370}

\bibitem[\protect\citeauthoryear{{Blanchette}, {Chayer}, {Wesemael},
  {Fontaine}, {Fontaine}, {Dupuis}, {Kruk}  \& {Green}}{{Blanchette}
  et~al.}{2008}]{Blanchette2008}
{Blanchette} J.-P.,  {Chayer} P.,  {Wesemael} F.,  {Fontaine} G.,  {Fontaine}
  M.,  {Dupuis} J.,  {Kruk} J.~W.,   {Green} E.~M.,  2008, \mn@doi [\apj]
  {10.1086/533580}, \href {http://adsabs.harvard.edu/abs/2008ApJ...678.1329B}
  {678, 1329}

\bibitem[\protect\citeauthoryear{{Brassard}, {Fontaine}, {Bill{\`e}res},
  {Charpinet}, {Liebert}  \& {Saffer}}{{Brassard} et~al.}{2001}]{Brassard2001}
{Brassard} P.,  {Fontaine} G.,  {Bill{\`e}res} M.,  {Charpinet} S.,  {Liebert}
  J.,   {Saffer} R.~A.,  2001, \mn@doi [\apj] {10.1086/323959}, \href
  {http://adsabs.harvard.edu/abs/2001ApJ...563.1013B} {563, 1013}

\bibitem[\protect\citeauthoryear{{Brooker} \& {Olle}}{{Brooker} \&
  {Olle}}{1955}]{Brooker1955}
{Brooker} R.~A.,  {Olle} T.~W.,  1955, \mn@doi [MNRAS]
  {10.1093/mnras/115.1.101}, \href
  {http://adsabs.harvard.edu/abs/1955MNRAS.115..101B} {115, 101}

\bibitem[\protect\citeauthoryear{{Brown}, {Ferguson}, {Davidsen}  \&
  {Dorman}}{{Brown} et~al.}{1997}]{Brown1997}
{Brown} T.~M.,  {Ferguson} H.~C.,  {Davidsen} A.~F.,   {Dorman} B.,  1997,
  \mn@doi [\apj] {10.1086/304187}, \href
  {http://adsabs.harvard.edu/abs/1997ApJ...482..685B} {482, 685}

\bibitem[\protect\citeauthoryear{{Bruntt}}{{Bruntt}}{2009}]{Bruntt2009}
{Bruntt} H.,  2009, \mn@doi [\aap] {10.1051/0004-6361/200911925}, \href
  {http://adsabs.harvard.edu/abs/2009A%26A...506..235B} {506, 235}

\bibitem[\protect\citeauthoryear{{Bruntt} et~al.,}{{Bruntt}
  et~al.}{2002}]{Bruntt2002}
{Bruntt} H.,  et~al., 2002, \mn@doi [\aap] {10.1051/0004-6361:20020549}, \href
  {http://adsabs.harvard.edu/abs/2002A%26A...389..345B} {389, 345}

\bibitem[\protect\citeauthoryear{{Bruntt} et~al.,}{{Bruntt}
  et~al.}{2004}]{Bruntt2004}
{Bruntt} H.,  et~al., 2004, \mn@doi [\aap] {10.1051/0004-6361:20040464}, \href
  {http://adsabs.harvard.edu/abs/2004A%26A...425..683B} {425, 683}

\bibitem[\protect\citeauthoryear{{Bruntt}, {De Cat}  \& {Aerts}}{{Bruntt}
  et~al.}{2008}]{Bruntt2008}
{Bruntt} H.,  {De Cat} P.,   {Aerts} C.,  2008, \mn@doi [\aap]
  {10.1051/0004-6361:20078523}, \href
  {http://adsabs.harvard.edu/abs/2008A%26A...478..487B} {478, 487}

\bibitem[\protect\citeauthoryear{{Bruntt} et~al.,}{{Bruntt}
  et~al.}{2010}]{Bruntt2010}
{Bruntt} H.,  et~al., 2010, \mn@doi [\mnras]
  {10.1111/j.1365-2966.2010.16575.x}, \href
  {http://adsabs.harvard.edu/abs/2010MNRAS.405.1907B} {405, 1907}

\bibitem[\protect\citeauthoryear{{Chen}, {Han}, {Deca}  \&
  {Podsiadlowski}}{{Chen} et~al.}{2013}]{Chen2013}
{Chen} X.,  {Han} Z.,  {Deca} J.,   {Podsiadlowski} P.,  2013, \mn@doi [\mnras]
  {10.1093/mnras/stt992}, \href
  {http://adsabs.harvard.edu/abs/2013MNRAS.434..186C} {434, 186}

\bibitem[\protect\citeauthoryear{{Claret} \& {Gimenez}}{{Claret} \&
  {Gimenez}}{1989}]{Claret1989}
{Claret} A.,  {Gimenez} A.,  1989, \aaps, \href
  {http://adsabs.harvard.edu/abs/1989A%26AS...81...37C} {81, 37}

\bibitem[\protect\citeauthoryear{{Copperwheat}, {Morales-Rueda}, {Marsh},
  {Maxted}  \& {Heber}}{{Copperwheat} et~al.}{2011}]{Copperwheat2011}
{Copperwheat} C.~M.,  {Morales-Rueda} L.,  {Marsh} T.~R.,  {Maxted} P.~F.~L.,
  {Heber} U.,  2011, \mn@doi [\mnras] {10.1111/j.1365-2966.2011.18786.x}, \href
  {http://adsabs.harvard.edu/abs/2011MNRAS.415.1381C} {415, 1381}

\bibitem[\protect\citeauthoryear{{Deca} et~al.,}{{Deca}
  et~al.}{2012}]{Deca2012}
{Deca} J.,  et~al., 2012, \mn@doi [\mnras] {10.1111/j.1365-2966.2012.20483.x},
  \href {http://adsabs.harvard.edu/abs/2012MNRAS.421.2798D} {421, 2798}

\bibitem[\protect\citeauthoryear{{Degroote} et~al.,}{{Degroote}
  et~al.}{2011}]{Degroote2011}
{Degroote} P.,  et~al., 2011, \mn@doi [\aap] {10.1051/0004-6361/201116802},
  \href {http://adsabs.harvard.edu/abs/2011A%26A...536A..82D} {536, A82}

\bibitem[\protect\citeauthoryear{{Demircan} \& {Kahraman}}{{Demircan} \&
  {Kahraman}}{1991}]{Demircan1991}
{Demircan} O.,  {Kahraman} G.,  1991, \mn@doi [\apss] {10.1007/BF00639097},
  \href {http://adsabs.harvard.edu/abs/1991Ap%26SS.181..313D} {181, 313}

\bibitem[\protect\citeauthoryear{{Edelmann}, {Heber}, {Hagen}, {Lemke},
  {Dreizler}, {Napiwotzki}  \& {Engels}}{{Edelmann}
  et~al.}{2003}]{Edelmann2003}
{Edelmann} H.,  {Heber} U.,  {Hagen} H.-J.,  {Lemke} M.,  {Dreizler} S.,
  {Napiwotzki} R.,   {Engels} D.,  2003, \mn@doi [\aap]
  {10.1051/0004-6361:20030135}, \href
  {http://adsabs.harvard.edu/abs/2003A%26A...400..939E} {400, 939}

\bibitem[\protect\citeauthoryear{{Freudling}, {Romaniello}, {Bramich},
  {Ballester}, {Forchi}, {Garc{\'{\i}}a-Dabl{\'o}}, {Moehler}  \&
  {Neeser}}{{Freudling} et~al.}{2013}]{reflex2013}
{Freudling} W.,  {Romaniello} M.,  {Bramich} D.~M.,  {Ballester} P.,  {Forchi}
  V.,  {Garc{\'{\i}}a-Dabl{\'o}} C.~E.,  {Moehler} S.,   {Neeser} M.~J.,  2013,
  \mn@doi [\aap] {10.1051/0004-6361/201322494}, \href
  {http://adsabs.harvard.edu/abs/2013A%26A...559A..96F} {559, A96}

\bibitem[\protect\citeauthoryear{{Geier}}{{Geier}}{2013}]{Geier2013}
{Geier} S.,  2013, \mn@doi [\aap] {10.1051/0004-6361/201220549}, \href
  {http://adsabs.harvard.edu/abs/2013A%26A...549A.110G} {549, A110}

\bibitem[\protect\citeauthoryear{{Geier}, {{\O}stensen}, {Nemeth}, {Gentile
  Fusillo}, {G{\"a}nsicke}, {Telting}, {Green}  \& {Schaffenroth}}{{Geier}
  et~al.}{2017}]{Geier2017}
{Geier} S.,  {{\O}stensen} R.~H.,  {Nemeth} P.,  {Gentile Fusillo} N.~P.,
  {G{\"a}nsicke} B.~T.,  {Telting} J.~H.,  {Green} E.~M.,   {Schaffenroth} J.,
  2017, \mn@doi [\aap] {10.1051/0004-6361/201630135}, \href
  {http://adsabs.harvard.edu/abs/2017A%26A...600A..50G} {600, A50}

\bibitem[\protect\citeauthoryear{{Green}, {Schmidt}  \& {Liebert}}{{Green}
  et~al.}{1986}]{Green1986}
{Green} R.~F.,  {Schmidt} M.,   {Liebert} J.,  1986, \mn@doi [\apjs]
  {10.1086/191115}, \href {http://adsabs.harvard.edu/abs/1986ApJS...61..305G}
  {61, 305}

\bibitem[\protect\citeauthoryear{{Green}, {Liebert}  \& {Saffer}}{{Green}
  et~al.}{2001}]{Green2001}
{Green} E.~M.,  {Liebert} J.,   {Saffer} R.~A.,  2001, in {J.~L.~Provencal,
  H.~L.~Shipman, J.~MacDonald, \& S.~Goodchild } ed.,  ASPCS Vol. 226, 12th
  European Workshop on White Dwarfs. p.~192 (\mn@eprint {}
  {arXiv:astro-ph/0012246})

\bibitem[\protect\citeauthoryear{{Green}, {Fontaine}, {Hyde}, {For}  \&
  {Chayer}}{{Green} et~al.}{2008}]{Green2008}
{Green} E.~M.,  {Fontaine} G.,  {Hyde} E.~A.,  {For} B.-Q.,   {Chayer} P.,
  2008, in {Heber} U.,  {Jeffery} C.~S.,   {Napiwotzki} R.,  eds,  Astronomical
  Society of the Pacific Conference Series Vol. 392, Hot Subdwarf Stars and
  Related Objects. p.~75

\bibitem[\protect\citeauthoryear{{Greggio} \& {Renzini}}{{Greggio} \&
  {Renzini}}{1990}]{Greggio1990}
{Greggio} L.,  {Renzini} A.,  1990, \mn@doi [\apj] {10.1086/169384}, \href
  {http://adsabs.harvard.edu/abs/1990ApJ...364...35G} {364, 35}

\bibitem[\protect\citeauthoryear{{Han}, {Tout}  \& {Eggleton}}{{Han}
  et~al.}{2000}]{Han2000}
{Han} Z.,  {Tout} C.~A.,   {Eggleton} P.~P.,  2000, \mn@doi [\mnras]
  {10.1046/j.1365-8711.2000.03839.x}, \href
  {http://adsabs.harvard.edu/abs/2000MNRAS.319..215H} {319, 215}

\bibitem[\protect\citeauthoryear{{Han}, {Podsiadlowski}, {Maxted}, {Marsh}  \&
  {Ivanova}}{{Han} et~al.}{2002}]{Han2002}
{Han} Z.,  {Podsiadlowski} P.,  {Maxted} P.~F.~L.,  {Marsh} T.~R.,   {Ivanova}
  N.,  2002, \mn@doi [\mnras] {10.1046/j.1365-8711.2002.05752.x}, \href
  {http://adsabs.harvard.edu/abs/2002MNRAS.336..449H} {336, 449}

\bibitem[\protect\citeauthoryear{{Han}, {Podsiadlowski}, {Maxted}  \&
  {Marsh}}{{Han} et~al.}{2003}]{Han2003}
{Han} Z.,  {Podsiadlowski} P.,  {Maxted} P.~F.~L.,   {Marsh} T.~R.,  2003,
  \mn@doi [\mnras] {10.1046/j.1365-8711.2003.06451.x}, \href
  {http://adsabs.harvard.edu/abs/2003MNRAS.341..669H} {341, 669}

\bibitem[\protect\citeauthoryear{{Heber}}{{Heber}}{1998}]{Heber1998}
{Heber} U.,  1998, in {Wamsteker} W.,  {Gonzalez Riestra} R.,   {Harris} B.,
  eds,  ESA Special Publication Vol. 413, Ultraviolet Astrophysics Beyond the
  IUE Final Archive. p.~195

\bibitem[\protect\citeauthoryear{{Heber}}{{Heber}}{2016}]{Heber2016}
{Heber} U.,  2016, \mn@doi [\pasp] {10.1088/1538-3873/128/966/082001}, \href
  {http://adsabs.harvard.edu/abs/2016PASP..128h2001H} {128, 082001}

\bibitem[\protect\citeauthoryear{{Heber}, {Moehler}, {Napiwotzki}, {Thejll}  \&
  {Green}}{{Heber} et~al.}{2002}]{Heber2002}
{Heber} U.,  {Moehler} S.,  {Napiwotzki} R.,  {Thejll} P.,   {Green} E.~M.,
  2002, \mn@doi [\aap] {10.1051/0004-6361:20020127}, \href
  {http://adsabs.harvard.edu/abs/2002A%26A...383..938H} {383, 938}

\bibitem[\protect\citeauthoryear{{Henden}, {Templeton}, {Terrell}, {Smith},
  {Levine}  \& {Welch}}{{Henden} et~al.}{2016}]{Henden2016}
{Henden} A.~A.,  {Templeton} M.,  {Terrell} D.,  {Smith} T.~C.,  {Levine} S.,
  {Welch} D.,  2016, VizieR Online Data Catalog, \href
  {http://cdsads.u-strasbg.fr/abs/2016yCat.2336....0H} {2336}

\bibitem[\protect\citeauthoryear{{Hubeny} \& {Lanz}}{{Hubeny} \&
  {Lanz}}{1995}]{Hubeny1995}
{Hubeny} I.,  {Lanz} T.,  1995, \mn@doi [\apj] {10.1086/175226}, \href
  {http://adsabs.harvard.edu/abs/1995ApJ...439..875H} {439, 875}

\bibitem[\protect\citeauthoryear{{H{\"u}mmerich}, {Bernhard}  \&
  {Srdoc}}{{H{\"u}mmerich} et~al.}{2014}]{Hummerich2014}
{H{\"u}mmerich} S.,  {Bernhard} K.,   {Srdoc} G.,  2014, Open European Journal
  on Variable Stars, \href {http://adsabs.harvard.edu/abs/2014OEJV..167....1H}
  {167, 1}

\bibitem[\protect\citeauthoryear{{Husser}, {Wende-von Berg}, {Dreizler},
  {Homeier}, {Reiners}, {Barman}  \& {Hauschildt}}{{Husser}
  et~al.}{2013}]{Husser2013}
{Husser} T.-O.,  {Wende-von Berg} S.,  {Dreizler} S.,  {Homeier} D.,  {Reiners}
  A.,  {Barman} T.,   {Hauschildt} P.~H.,  2013, \mn@doi [\aap]
  {10.1051/0004-6361/201219058}, \href
  {http://adsabs.harvard.edu/abs/2013A%26A...553A...6H} {553, A6}

\bibitem[\protect\citeauthoryear{{Kawka}, {Vennes}, {O'Toole}, {N{\'e}meth},
  {Burton}, {Kotze}  \& {Buckley}}{{Kawka} et~al.}{2015}]{Kawka2015}
{Kawka} A.,  {Vennes} S.,  {O'Toole} S.,  {N{\'e}meth} P.,  {Burton} D.,
  {Kotze} E.,   {Buckley} D.~A.~H.,  2015, \mn@doi [\mnras]
  {10.1093/mnras/stv821}, \href
  {http://adsabs.harvard.edu/abs/2015MNRAS.450.3514K} {450, 3514}

\bibitem[\protect\citeauthoryear{{Kilkenny}, {Heber}  \& {Drilling}}{{Kilkenny}
  et~al.}{1988}]{Kilkenny1988}
{Kilkenny} D.,  {Heber} U.,   {Drilling} J.~S.,  1988, South African
  Astronomical Observatory Circular, \href
  {http://cdsads.u-strasbg.fr/abs/1988SAAOC..12....1K} {12}

\bibitem[\protect\citeauthoryear{{Koen}, {Orosz}  \& {Wade}}{{Koen}
  et~al.}{1998}]{Koen1998}
{Koen} C.,  {Orosz} J.~A.,   {Wade} R.~A.,  1998, \mn@doi [\mnras]
  {10.1046/j.1365-8711.1998.01961.x}, \href
  {http://adsabs.harvard.edu/abs/1998MNRAS.300..695K} {300, 695}

\bibitem[\protect\citeauthoryear{{Kupfer} et~al.,}{{Kupfer}
  et~al.}{2015}]{Kupfer2015}
{Kupfer} T.,  et~al., 2015, VizieR Online Data Catalog, \href
  {http://adsabs.harvard.edu/abs/2015yCat..35760044K} {357}

\bibitem[\protect\citeauthoryear{{Lamontagne}, {Demers}, {Wesemael}, {Fontaine}
   \& {Irwin}}{{Lamontagne} et~al.}{2000}]{Lamontagne2000}
{Lamontagne} R.,  {Demers} S.,  {Wesemael} F.,  {Fontaine} G.,   {Irwin} M.~J.,
   2000, \mn@doi [\aj] {10.1086/301181}, \href
  {http://cdsads.u-strasbg.fr/abs/2000AJ....119..241L} {119, 241}

\bibitem[\protect\citeauthoryear{{Lubow} \& {Shu}}{{Lubow} \&
  {Shu}}{1975}]{Lubow1975}
{Lubow} S.~H.,  {Shu} F.~H.,  1975, \mn@doi [\apj] {10.1086/153614}, \href
  {http://adsabs.harvard.edu/abs/1975ApJ...198..383L} {198, 383}

\bibitem[\protect\citeauthoryear{{Luo}, {Nemeth}, {Liu}, {Deng}  \&
  {Han}}{{Luo} et~al.}{2016}]{Luo2016}
{Luo} Y.,  {Nemeth} P.,  {Liu} C.,  {Deng} L.,   {Han} Z.,  2016, preprint,
  \href {http://adsabs.harvard.edu/abs/2016arXiv160103507L} {} (\mn@eprint
  {arXiv} {1601.03507})

\bibitem[\protect\citeauthoryear{{Martin} et~al.,}{{Martin}
  et~al.}{2005}]{Martin2005}
{Martin} D.~C.,  et~al., 2005, \mn@doi [\apjl] {10.1086/426387}, \href
  {http://adsabs.harvard.edu/abs/2005ApJ...619L...1M} {619, L1}

\bibitem[\protect\citeauthoryear{{Maxted}, {Moran}, {Marsh}  \&
  {Gatti}}{{Maxted} et~al.}{2000}]{Maxted2000}
{Maxted} P.~F.~L.,  {Moran} C.~K.~J.,  {Marsh} T.~R.,   {Gatti} A.~A.,  2000,
  \mn@doi [\mnras] {10.1046/j.1365-8711.2000.03102.x}, \href
  {http://adsabs.harvard.edu/abs/2000MNRAS.311..877M} {311, 877}

\bibitem[\protect\citeauthoryear{{Maxted}, {Heber}, {Marsh}  \&
  {North}}{{Maxted} et~al.}{2001}]{Maxted2001}
{Maxted} P.~f.~L.,  {Heber} U.,  {Marsh} T.~R.,   {North} R.~C.,  2001, \mn@doi
  [\mnras] {10.1046/j.1365-8711.2001.04714.x}, \href
  {http://adsabs.harvard.edu/abs/2001MNRAS.326.1391M} {326, 1391}

\bibitem[\protect\citeauthoryear{{Moehler}, {Richtler}, {de Boer}, {Dettmar}
  \& {Heber}}{{Moehler} et~al.}{1990}]{Moehler1990}
{Moehler} S.,  {Richtler} T.,  {de Boer} K.~S.,  {Dettmar} R.~J.,   {Heber} U.,
   1990, \aaps, \href {http://adsabs.harvard.edu/abs/1990A%26AS...86...53M}
  {86, 53}

\bibitem[\protect\citeauthoryear{{Morales-Rueda}, {Maxted}, {Marsh}, {North}
  \& {Heber}}{{Morales-Rueda} et~al.}{2003}]{Morales2003}
{Morales-Rueda} L.,  {Maxted} P.~F.~L.,  {Marsh} T.~R.,  {North} R.~C.,
  {Heber} U.,  2003, \mn@doi [\mnras] {10.1046/j.1365-8711.2003.06088.x}, \href
  {http://adsabs.harvard.edu/abs/2003MNRAS.338..752M} {338, 752}

\bibitem[\protect\citeauthoryear{{Napiwotzki}, {Karl}, {Lisker}, {Heber},
  {Christlieb}, {Reimers}, {Nelemans}  \& {Homeier}}{{Napiwotzki}
  et~al.}{2004}]{Napiwotzki2004}
{Napiwotzki} R.,  {Karl} C.~A.,  {Lisker} T.,  {Heber} U.,  {Christlieb} N.,
  {Reimers} D.,  {Nelemans} G.,   {Homeier} D.,  2004, \mn@doi [\apss]
  {10.1023/B:ASTR.0000044362.07416.6c}, \href
  {http://adsabs.harvard.edu/abs/2004Ap%26SS.291..321N} {291, 321}

\bibitem[\protect\citeauthoryear{{N{\'e}meth}, {Kawka}  \&
  {Vennes}}{{N{\'e}meth} et~al.}{2012}]{Nemeth2012}
{N{\'e}meth} P.,  {Kawka} A.,   {Vennes} S.,  2012, \mn@doi [\mnras]
  {10.1111/j.1365-2966.2012.22009.x}, \href
  {http://adsabs.harvard.edu/abs/2012MNRAS.427.2180N} {427, 2180}

\bibitem[\protect\citeauthoryear{{O'Toole} \& {Heber}}{{O'Toole} \&
  {Heber}}{2006}]{Otoole2006}
{O'Toole} S.~J.,  {Heber} U.,  2006, \mn@doi [\aap]
  {10.1051/0004-6361:20053948}, \href
  {http://adsabs.harvard.edu/abs/2006A%26A...452..579O} {452, 579}

\bibitem[\protect\citeauthoryear{{{\O}stensen} \& {Van Winckel}}{{{\O}stensen}
  \& {Van Winckel}}{2011}]{Oestensen2011}
{{\O}stensen} R.~H.,  {Van Winckel} H.,  2011, in {L.~Schmidtobreick,
  M.~R.~Schreiber, \& C.~Tappert} ed.,  ASPCS Vol. 447, Evolution of Compact
  Binaries. p.~171

\bibitem[\protect\citeauthoryear{{{\O}stensen} \& {Van Winckel}}{{{\O}stensen}
  \& {Van Winckel}}{2012}]{Oestensen2012}
{{\O}stensen} R.~H.,  {Van Winckel} H.,  2012, in {D.~Kilkenny, C.~S.~Jeffery,
  \& C.~Koen} ed.,  ASPCS Vol. 452, Fifth Meeting on Hot Subdwarf Stars and
  Related Objects. p.~163 (\mn@eprint {arXiv} {1112.0977})

\bibitem[\protect\citeauthoryear{{Paczynski}}{{Paczynski}}{1976}]{Paczynski1976}
{Paczynski} B.,  1976, in {Eggleton} P.,  {Mitton} S.,   {Whelan} J.,  eds,
  IAU Symposium Vol. 73, Structure and Evolution of Close Binary Systems. p.~75

\bibitem[\protect\citeauthoryear{{Pecaut} \& {Mamajek}}{{Pecaut} \&
  {Mamajek}}{2013}]{Pecaut2013}
{Pecaut} M.~J.,  {Mamajek} E.~E.,  2013, \mn@doi [\apjs]
  {10.1088/0067-0049/208/1/9}, \href
  {http://adsabs.harvard.edu/abs/2013ApJS..208....9P} {208, 9}

\bibitem[\protect\citeauthoryear{{Reed} \& {Stiening}}{{Reed} \&
  {Stiening}}{2004}]{Reed2004}
{Reed} M.~D.,  {Stiening} R.,  2004, \mn@doi [\pasp] {10.1086/421253}, \href
  {http://cdsads.u-strasbg.fr/abs/2004PASP..116..506R} {116, 506}

\bibitem[\protect\citeauthoryear{{Ryabchikova}, {Piskunov}, {Kurucz},
  {Stempels}, {Heiter}, {Pakhomov}  \& {Barklem}}{{Ryabchikova}
  et~al.}{2015}]{Ryabchikova2015}
{Ryabchikova} T.,  {Piskunov} N.,  {Kurucz} R.~L.,  {Stempels} H.~C.,  {Heiter}
  U.,  {Pakhomov} Y.,   {Barklem} P.~S.,  2015, \mn@doi [\physscr]
  {10.1088/0031-8949/90/5/054005}, \href
  {http://adsabs.harvard.edu/abs/2015PhyS...90e4005R} {90, 054005}

\bibitem[\protect\citeauthoryear{{Saffer}, {Bergeron}, {Koester}  \&
  {Liebert}}{{Saffer} et~al.}{1994}]{Saffer1994}
{Saffer} R.~A.,  {Bergeron} P.,  {Koester} D.,   {Liebert} J.,  1994, \mn@doi
  [\apj] {10.1086/174573}, \href
  {http://adsabs.harvard.edu/abs/1994ApJ...432..351S} {432, 351}

\bibitem[\protect\citeauthoryear{{Shulyak}, {Tsymbal}, {Ryabchikova},
  {St{\"u}tz}  \& {Weiss}}{{Shulyak} et~al.}{2004}]{Shulyak2004}
{Shulyak} D.,  {Tsymbal} V.,  {Ryabchikova} T.,  {St{\"u}tz} C.,   {Weiss}
  W.~W.,  2004, \mn@doi [\aap] {10.1051/0004-6361:20034169}, \href
  {http://adsabs.harvard.edu/abs/2004A%26A...428..993S} {428, 993}

\bibitem[\protect\citeauthoryear{{Skrutskie} et~al.,}{{Skrutskie}
  et~al.}{2006}]{Skrutskie2006}
{Skrutskie} M.~F.,  et~al., 2006, \mn@doi [\aj] {10.1086/498708}, \href
  {http://adsabs.harvard.edu/abs/2006AJ....131.1163S} {131, 1163}

\bibitem[\protect\citeauthoryear{{Stobie} et~al.,}{{Stobie}
  et~al.}{1997}]{Stobie1997}
{Stobie} R.~S.,  et~al., 1997, \mn@doi [\mnras] {10.1093/mnras/287.4.848},
  \href {http://adsabs.harvard.edu/abs/1997MNRAS.287..848S} {287, 848}

\bibitem[\protect\citeauthoryear{{Tauris} \& {van den Heuvel}}{{Tauris} \& {van
  den Heuvel}}{2006}]{Tauris2006}
{Tauris} T.~M.,  {van den Heuvel} E.~P.~J.,  2006, {Formation and evolution of
  compact stellar X-ray sources}.
pp 623--665

\bibitem[\protect\citeauthoryear{{Tkachenko}}{{Tkachenko}}{2015}]{Tkachenko2015}
{Tkachenko} A.,  2015, \mn@doi [\aap] {10.1051/0004-6361/201526513}, \href
  {http://adsabs.harvard.edu/abs/2015A%26A...581A.129T} {581, A129}

\bibitem[\protect\citeauthoryear{{Tsymbal}}{{Tsymbal}}{1996}]{Tsymbal1996}
{Tsymbal} V.,  1996, in {Adelman} S.~J.,  {Kupka} F.,   {Weiss} W.~W.,  eds,
  Astronomical Society of the Pacific Conference Series Vol. 108, M.A.S.S.,
  Model Atmospheres and Spectrum Synthesis. p.~198

\bibitem[\protect\citeauthoryear{{Ulrich} \& {Burger}}{{Ulrich} \&
  {Burger}}{1976}]{Ulrich1976}
{Ulrich} R.~K.,  {Burger} H.~L.,  1976, \mn@doi [\apj] {10.1086/154406}, \href
  {http://adsabs.harvard.edu/abs/1976ApJ...206..509U} {206, 509}

\bibitem[\protect\citeauthoryear{{Vennes}, {Kawka}  \& {N{\'e}meth}}{{Vennes}
  et~al.}{2011}]{Vennes2011}
{Vennes} S.,  {Kawka} A.,   {N{\'e}meth} P.,  2011, \mn@doi [\mnras]
  {10.1111/j.1365-2966.2010.17584.x}, \href
  {http://adsabs.harvard.edu/abs/2011MNRAS.410.2095V} {410, 2095}

\bibitem[\protect\citeauthoryear{{Vos} et~al.,}{{Vos} et~al.}{2012}]{Vos2012}
{Vos} J.,  et~al., 2012, \mn@doi [\aap] {10.1051/0004-6361/201219723}, \href
  {http://adsabs.harvard.edu/abs/2012A%26A...548A...6V} {548, A6}

\bibitem[\protect\citeauthoryear{{Vos}, {{\O}stensen}, {N{\'e}meth}, {Green},
  {Heber}  \& {Van Winckel}}{{Vos} et~al.}{2013}]{Vos2013}
{Vos} J.,  {{\O}stensen} R.~H.,  {N{\'e}meth} P.,  {Green} E.~M.,  {Heber} U.,
   {Van Winckel} H.,  2013, \mn@doi [\aap] {10.1051/0004-6361/201322200}, \href
  {http://adsabs.harvard.edu/abs/2013A%26A...559A..54V} {559, A54}

\bibitem[\protect\citeauthoryear{{Vos}, {{\O}stensen}  \& {Van Winckel}}{{Vos}
  et~al.}{2014}]{Vos2014}
{Vos} J.,  {{\O}stensen} R.,   {Van Winckel} H.,  2014, in {van Grootel} V.,
  {Green} E.,  {Fontaine} G.,   {Charpinet} S.,  eds,  Astronomical Society of
  the Pacific Conference Series Vol. 481, 6th Meeting on Hot Subdwarf Stars and
  Related Objects. p.~265 (\mn@eprint {arXiv} {1310.0189})

\bibitem[\protect\citeauthoryear{{Vos}, {{\O}stensen}, {Marchant}  \& {Van
  Winckel}}{{Vos} et~al.}{2015}]{Vos2015}
{Vos} J.,  {{\O}stensen} R.~H.,  {Marchant} P.,   {Van Winckel} H.,  2015,
  \mn@doi [\aap] {10.1051/0004-6361/201526019}, \href
  {http://adsabs.harvard.edu/abs/2015A%26A...579A..49V} {579, A49}

\bibitem[\protect\citeauthoryear{{Vos}, {{\O}stensen}, {Vu\u{c}kovi\'{c}}  \&
  {Van Winckel}}{{Vos} et~al.}{2017}]{Vos2017}
{Vos} J.,  {{\O}stensen} R.~H.,  {Vu\u{c}kovi\'{c}} M.,   {Van Winckel} H.,
  2017, \aap\ in Press, \href
  {http://adsabs.harvard.edu/abs/2017arXiv170602286V} {}

\bibitem[\protect\citeauthoryear{{Webbink}}{{Webbink}}{1984}]{Webbink1984}
{Webbink} R.~F.,  1984, \mn@doi [\apj] {10.1086/161701}, \href
  {http://adsabs.harvard.edu/abs/1984ApJ...277..355W} {277, 355}

\bibitem[\protect\citeauthoryear{{Zahn}}{{Zahn}}{1977}]{Zahn1977}
{Zahn} J.-P.,  1977, A\&A, \href
  {http://adsabs.harvard.edu/abs/1977A%26A....57..383Z} {57, 383}

\makeatother
\end{thebibliography}

\bsp	
\label{lastpage}
\end{document}